\documentclass[12pt,letterpaper]{article}
\usepackage{epsfig,rotating,setspace,latexsym,amsmath,epsf,amssymb,bm}
\usepackage{cite}

\title{Dependence Balance Based Outer Bounds for Gaussian Networks with Cooperation and Feedback\thanks{This work was supported by
NSF Grants CCF $04$-$47613$, CCF $05$-$14846$, CNS $07$-$16311$
and CCF 07-29127. }}

\author{Ravi Tandon \qquad Sennur Ulukus \\
\normalsize Department of Electrical and Computer Engineering\\
\normalsize University of Maryland, College Park, MD 20742 \\
\normalsize {\it ravit@umd.edu} \qquad {\it ulukus@umd.edu}}

\newtheorem{Theo}{Theorem}
\newtheorem{Lem}{Lemma}

\begin{document}
\date{}
\maketitle
\thispagestyle{empty} \vspace{-0.39in}
\begin{abstract}
We obtain new outer bounds on the capacity regions of the two-user
multiple access channel with generalized feedback (MAC-GF) and the two-user
interference channel with generalized feedback (IC-GF). These
outer bounds are based on the idea of dependence balance which was
proposed by Hekstra and Willems \cite{Hekstra_Willems:1989}. To
illustrate the usefulness of our outer bounds, we investigate three
different channel models.

We first consider a Gaussian MAC with noisy feedback (MAC-NF), where transmitter $k$,
$k=1,2$, receives a feedback $Y_{F_{k}}$, which is the channel output
$Y$ corrupted with additive white Gaussian noise $Z_{k}$. As the
feedback noise variances $\sigma_{Z_{k}}^{2}$, $k=1,2$, become
large, one would expect the feedback to become useless. This fact is not reflected by the cut-set outer bound.
We demonstrate that our outer bound improves upon the cut-set bound for all non-zero
values of the feedback noise variances. Moreover, in the limit as
$\sigma_{Z_{k}}^{2}\rightarrow \infty$, $k=1,2$, our outer bound
collapses to the capacity region of the Gaussian MAC without
feedback.

Secondly, we investigate a Gaussian MAC with user-cooperation
(MAC-UC), where each transmitter receives an additive white
Gaussian noise corrupted version of the channel input of the other
transmitter \cite{Aazhang:2003}.
For this channel model, the
cut-set bound is sensitive to the cooperation noises, but not
sensitive enough. For all non-zero
values of cooperation noise variances, our outer bound strictly
improves upon the cut-set outer bound.
Moreover, as the cooperation noises become large, our outer bound collapses to the capacity
region of the Gaussian MAC without cooperation.

Thirdly, we investigate a Gaussian IC with user-cooperation
(IC-UC). For this channel model, the cut-set bound is again
sensitive to cooperation noise variances as in the case of MAC-UC
channel model, but not sensitive enough. We demonstrate that our
outer bound strictly improves upon the cut-set bound for all
non-zero values of cooperation noise variances.
\end{abstract}

\section{Introduction}
It is well known that noiseless feedback can increase the
capacity region of the discrete memoryless multiple access channel
as was shown by Gaarder and Wolf in \cite{GW:1975}. The multiple access
channel with generalized feedback (MAC-GF) was first introduced by
Carleial \cite{Carleial:1982}. The model therein allows for different
feedback signals at the two transmitters. For this channel model,
Carleial \cite{Carleial:1982} obtained an achievable rate region
using block Markov superposition encoding and windowed decoding.
An improvement  over this achievable rate region was obtained by
Willems et. al. in \cite{WVMS:1983} by using block Markov
superposition encoding combined with backwards decoding.

Inspired from the uplink MAC-GF channel model, the interference
channel with generalized feedback (IC-GF) was studied in
\cite{Tuninetti:2007}, \cite{Host-Madsen:2006}, (also see the
references therein) where achievable rate regions were obtained.
It was shown in \cite{Tuninetti:2007} and \cite{Host-Madsen:2006}
that for the Gaussian interference channel with user cooperation
(IC-UC), the overheard information at the transmitters has a dual
effect of enabling cooperation and mitigating interference,
thereby providing improved achievable rates compared to the best
known evaluation of the Han-Kobayashi achievable rate region
\cite{Han:1981}, \cite{Sason:2004}.

As far as the converses are concerned for the MAC-GF and the
IC-GF, a well known outer bound is the cut-set outer bound. The
cut-set bound allows all input distributions, thereby permitting
arbitrary correlation between the channel inputs and hence is
seemingly loose. The idea of dependence balance was first
introduced by Hekstra and Willems \cite{Hekstra_Willems:1989} to
obtain outer bounds on the capacity region of single output
two-way channel. In contrast to the cut-set bound, the dependence
balance bound provides an additional non-trivial restriction over
the set of allowable input distributions thus leading to a
potentially tighter outer bound. In the same paper
\cite{Hekstra_Willems:1989}, the authors give a variant of this
bound for the two-user discrete memoryless MAC with noiseless
feedback from the receiver.

In this paper, we use the idea of dependence balance to obtain new
outer bounds on the capacity regions of the MAC-GF and the IC-GF.
To show the usefulness of our outer bounds, we will consider three
different channel models.

We first consider the Gaussian MAC with different noisy feedback
signals at the two transmitters. Specifically, transmitter $k$,
$k=1,2$, receives a feedback $Y_{F_{k}}=Y+Z_{k}$, where $Y$ is the
received signal and $Z_{k}$ is zero-mean, Gaussian random variable
with variance $\sigma_{Z_{k}}^{2}$. The capacity region is only
known when feedback is noiseless, i.e., $Y_{F_{1}}=Y_{F_{2}}=Y$,
in which case the feedback capacity region equals the cut-set
outer bound, as was shown by Ozarow \cite{OZ:1984}. For the case
of noisy feedback in consideration, the cut-set outer bound is
insensitive to the noise in feedback links, i.e., it is not
sensitive to the variances of $Z_{1}$ and $Z_{2}$. As the feedback
becomes more corrupted, or in other words, as
$\sigma_{Z_{1}}^{2},\sigma_{Z_{2}}^{2}$ become large, one would
expect the feedback to become useless. This fact is not accounted
for by the cut-set bound. We show that our outer bound strictly
improves upon the cut-set bound for all non-zero values of
$(\sigma_{Z_{1}}^{2},\sigma_{Z_{2}}^{2})$. Furthermore, as
$(\sigma_{Z_{1}}^{2},\sigma_{Z_{2}}^{2})$ become large, our outer
bound collapses to the capacity region of Gaussian MAC without
feedback, thereby establishing the feedback capacity region. We
should mention here that applying the idea of dependence balance
to obtain improved outer bounds for Gaussian MAC with noisy
feedback was proposed by Gastpar and Kramer in \cite{GK1:2006}.

Secondly, we investigate the Gaussian MAC with transmitter
cooperation. Sendonaris, Erkip and Aazhang \cite{Aazhang:2003}
studied a model where each transmitter receives a version of the
other transmitter's current channel input corrupted with additive
white Gaussian noise. They named this model as \emph{user
cooperation} model. This model is particularly suitable for a
wireless setting since the transmitters can potentially overhear
each other. An achievable rate region for the user cooperation
model was given in \cite{Aazhang:2003} using the result of
\cite{WVMS:1983} and was shown to strictly exceed the rate region
if the transmitters ignore the overheard signals.

We evaluate our outer bound for the user cooperation setting
described above. In contrast to the case of noisy feedback, the
cut-set bound for the user cooperation model is sensitive to
cooperation noise variances, but not too sensitive. Intuitively
speaking, as the backward noise variances become large, one would
expect the cut-set bound to collapse to the capacity region of the
MAC without feedback. Instead, the cut-set bound converges to the
capacity region of the Gaussian MAC with noiseless output feedback
\cite{OZ:1984}. On the other hand, in the limit when cooperation
noise variances become too large, our bound converges to the
capacity region of the Gaussian MAC with no cooperation, thereby
yielding a capacity result. For all non-zero and finite values of
cooperation noise variances, our outer bound strictly improves
upon the cut-set outer bound. Our dependence balance based outer
bound coincides with the cut-set bound only when the backward
noise variance is identically zero and both outer bounds collapse
to the total cooperation line.

Thirdly, we evaluate our outer bound for the Gaussian IC with user
cooperation (IC-UC). For all non-zero and finite values of
cooperation noise variances, our outer bound strictly improves
upon the cut-set outer bound. We should remark here that the
approach of dependence balance was also used in \cite{GK3:2006} to
obtain an improved sum-rate upper bound for the Gaussian IC with
common, noisy feedback from the receivers.

Evaluation of our outer bounds for MAC-NF, MAC-UC and IC-UC is not
straightforward since our outer bounds are expressed in terms of a
union of probability densities of three random variables, one of
which is an auxiliary random variable. Moreover, these unions are
over all such densities which satisfy a non-trivial dependence
balance constraint. We overcome this difficulty by proving
separately for all three models in consideration, that it is
sufficient to consider jointly Gaussian input distributions,
satisfying the dependence balance constraint, when evaluating our
outer bounds. The proof methodology for showing this claim is
entirely different for each of the cases of noisy feedback and
user cooperation models. In particular, for the case of MAC-NF, we
make use of a recently discovered multivariate generalization
\cite{Palomar:2008} of Costa's entropy power inequality (EPI)
\cite{Costa:EPI1985} along with some properties of $3 \times 3$
covariance matrices to obtain this result. On the other hand, for
the case of MAC-UC and IC-UC, we do not need EPI to show this
result and our proof closely follows the proof of a recent result
by Bross, Lapidoth and Wigger
\cite{LapidothISIT:2008},\cite{Venkat:thesis} for the Gaussian MAC
with conferencing encoders. The structure of dependence balance
constraints for the channel models in consideration are of
different form, which explains the different methodology of
proofs.

For the most general setting of MAC-GF and IC-GF, our outer bounds are
expressed in terms of two auxiliary random variables. For the three
channel models in consideration, i.e., MAC-NF, MAC-UC and IC-UC, we suitably modify
our outer bounds to express them in terms of only one auxiliary random variable.
These modifications are particularly helpful in their explicit evaluation.
We also believe that the proof methodology developed for
evaluating our outer bounds could be helpful for other multi-user
information theoretic problems.

\section{System Model}
\subsection{MAC with Generalized Feedback}

A discrete memoryless two-user multiple access channel with
generalized feedback (MAC-GF) (see Figure $1$) is defined by: two
input alphabets $\mathcal{X}_{1}$ and $\mathcal{X}_{2}$, an output
alphabet for the receiver $\mathcal{Y}$, feedback output alphabets
$\mathcal{Y}_{F_{1}}$ and $\mathcal{Y}_{F_{2}}$ at transmitters $1$ and
$2$, respectively, and a probability transition function
$p(y,y_{F_{1}},y_{F_{2}}|x_{1},x_{2})$, defined for all triples
$(y,y_{F_{1}},y_{F_{2}})\in \mathcal{Y}\times\mathcal{Y}_{F_{1}}\times \mathcal{Y}_{F_{2}}$, for every
pair $(x_{1},x_{2})\in\mathcal{X}_{1}\times\mathcal{X}_{2}$.

A $(n,M_{1},M_{2},P_{e})$ code for the MAC-GF consists of two sets
of encoding functions
$f_{1i}:\mathcal{M}_{1}\times\mathcal{Y}_{F_{1}}^{i-1}\rightarrow
\mathcal{X}_{1}$,
$f_{2i}:\mathcal{M}_{2}\times\mathcal{Y}_{F_{2}}^{i-1}\rightarrow
\mathcal{X}_{2}$ for $i=1,\ldots,n$ and a decoding function $g:
\mathcal{Y}^{n} \rightarrow \mathcal{M}_{1} \times
\mathcal{M}_{2}$. The two transmitters produce independent and
uniformly distributed messages $W_{1} \in \{1,\ldots,M_{1}\}$ and
$W_{2} \in \{1,\ldots,M_{2}\}$, respectively, and transmit them
through $n$ channel uses. The average error probability is defined
as, $P_{e}=\mbox{Pr}[(\hat{W}_{1},\hat{W}_{2})\neq
(W_{1},W_{2})]$. A rate pair $(R_{1},R_{2})$ is said to be
achievable for MAC-GF if for any $\epsilon \geq 0$, there exists a
pair of $n$ encoding functions $\{f_{1i}\}_{i=1}^{n}$,
$\{f_{2i}\}_{i=1}^{n}$, and a decoding function
$g:\mathcal{Y}^{n}\rightarrow \mathcal{M}_{1} \times
\mathcal{M}_{2}$ such that $ R_{1}\leq \text{log}(M_{1})/n$,
$R_{2}\leq \text{log}(M_{2})/n$ and $P_{e}\leq \epsilon$ for
sufficiently large $n$. The capacity region of MAC-GF is the
closure of the set of all achievable rate pairs $(R_{1},R_{2})$.

\subsection{IC with Generalized Feedback}

A discrete memoryless two-user interference channel with
generalized feedback (IC-GF) (see Figure $2$) is defined by: two
input alphabets $\mathcal{X}_{1}$ and $\mathcal{X}_{2}$, two
output alphabets $\mathcal{Y}_{1}$ and $\mathcal{Y}_{2}$ at
receivers $1$ and $2$, respectively, two feedback output alphabets
$\mathcal{Y}_{F_{1}}$ and $\mathcal{Y}_{F_{2}}$ at transmitters
$1$ and $2$, respectively, and a probability transition function
$p(y_{1},y_{2},y_{F_{1}},y_{F_{2}}|x_{1},x_{2})$, defined for all
quadruples $(y_{1},y_{2},y_{F_{1}},y_{F_{2}})\in
\mathcal{Y}_{1}\times
\mathcal{Y}_{2}\times\mathcal{Y}_{F_{1}}\times\mathcal{Y}_{F_{2}}$,
for every pair
$(x_{1},x_{2})\in\mathcal{X}_{1}\times\mathcal{X}_{2}$.

A $(n,M_{1},M_{2},P_{e}^{(1)},P_{e}^{(2)})$ code for IC-GF
consists of two sets of encoding functions
$f_{1i}:\mathcal{M}_{1}\times\mathcal{Y}_{F_{1}}^{i-1}\rightarrow
\mathcal{X}_{1}$,
$f_{2i}:\mathcal{M}_{2}\times\mathcal{Y}_{F_{2}}^{i-1}\rightarrow
\mathcal{X}_{2}$ for $i=1,\ldots,n$ and two decoding functions
$g_{1}: \mathcal{Y}_{1}^{n} \rightarrow
\mathcal{M}_{1}$ and $g_{2}: \mathcal{Y}_{2}^{n} \rightarrow
\mathcal{M}_{2}$. The two transmitters produce
independent and uniformly distributed messages $W_{1} \in
\{1,\ldots,M_{1}\}$ and $W_{2} \in \{1,\ldots,M_{2}\}$,
respectively, and transmit them through $n$ channel uses. The
average error probability at receivers $1$ and $2$ are defined as,
$P_{e}^{(k)}= \mbox{Pr}[\hat{W}_{k}\neq
W_{k}]$ for  $k=1,2$.
A rate pair $(R_{1},R_{2})$ is said to be achievable for
IC-GF if for any pair $\epsilon_{1} \geq 0, \epsilon_{2} \geq 0$, there exists a pair of  $n$
encoding functions $\{f_{1i}\}_{i=1}^{n}$, $\{f_{2i}\}_{i=1}^{n}$,
and a pair of decoding functions $(g_{1},g_{2})$ such that $R_{1}\leq
\text{log}(M_{1})/n$, $R_{2}\leq \text{log}(M_{2})/n$ and $P_{e}^{(k)}\leq \epsilon_{k}$ for
sufficiently large $n$, for $k=1,2$. The capacity region of IC-GF is the closure of the set of all achievable rate
pairs $(R_{1},R_{2})$.
\begin{figure}[t]
  \centerline{\epsfig{figure=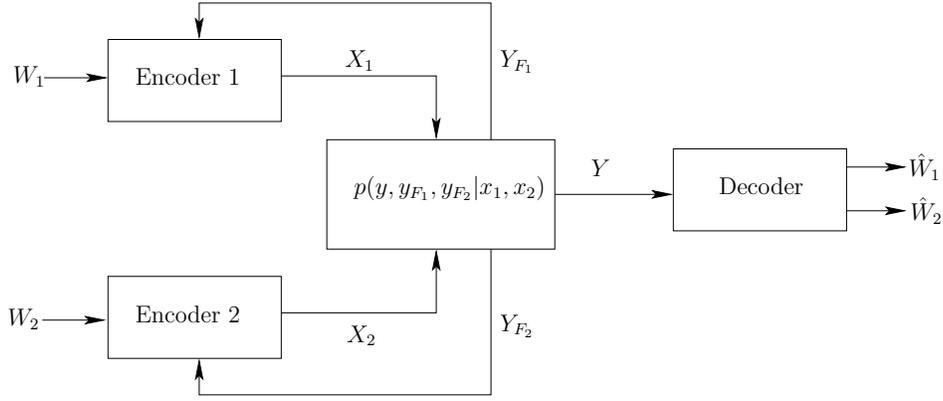,width=12.5cm}}
  \caption{The multiple access channel with generalized feedback (MAC-GF).}\label{fig1}
\end{figure}

\section{Cut-set Outer Bounds}
A general outer bound on the capacity region of a multi-terminal network is the cut-set outer bound \cite{Cover:book}.
The cut-set outer bound for MAC-GF is given by
\begin{align}
\mathcal{CS}^{MAC}=\big\{(R_{1},R_{2}):\hspace{0.05in}&R_{1}\leq I(X_{1};Y,Y_{F_{2}}|X_{2})\label{CSMAC1}\\
&R_{2}\leq I(X_{2};Y,Y_{F_{1}}|X_{1})\label{CSMAC2}\\
&R_{1}+R_{2}\leq I(X_{1},X_{2};Y)\big\}\label{CSMAC3}
\end{align}
where the random variables $X_{1},X_{2}$ and $(Y,Y_{F_{1}},Y_{F_{2}})$
have the joint distribution
\begin{align}
&\hspace{0.3in}p(x_{1},x_{2},y,y_{F_{1}},y_{F_{2}})=p(x_{1},x_{2})p(y,y_{F_{1}},y_{F_{2}}|x_{1},x_{2}).\label{CSMAC4}
\end{align}
The cut-set outer bound for IC-GF is
given by
\begin{align}
\mathcal{CS}^{IC}=\big\{(R_{1},R_{2}):\hspace{0.05in}&R_{1}\leq I(X_{1},X_{2};Y_{1})\label{CSIC1}\\
&R_{2}\leq I(X_{1},X_{2};Y_{2})\label{CSIC2}\\
&R_{1}\leq I(X_{1};Y_{1},Y_{2},Y_{F_{2}}|X_{2})\label{CSIC3}\\
&R_{2}\leq I(X_{2};Y_{1},Y_{2},Y_{F_{1}}|X_{1})\label{CSIC4}\\
&R_{1}+R_{2}\leq I(X_{1},X_{2};Y_{1},Y_{2})\big\}\label{CSIC5}
\end{align}
where the random variables $X_{1},X_{2}$ and $(Y_{1},Y_{2},Y_{F_{1}},Y_{F_{2}})$ have the joint
distribution
\begin{align}
&\hspace{0.3in}p(x_{1},x_{2},y_{1},y_{2},y_{F_{1}},y_{F_{2}})=p(x_{1},x_{2})p(y_{1},y_{2},y_{F_{1}},y_{F_{2}}|x_{1},x_{2}).\label{CSIC6}
\end{align}
The cut-set bound is seemingly loose since it allows arbitrary correlation among channel inputs by
permitting arbitrary input distributions $p(x_{1},x_{2})$. Using the approach of dependence balance, we will obtain
outer bounds for MAC-GF and IC-GF which restrict the corresponding set of input distributions for both channel models.
In particular, our outer bounds only permit those input distributions which satisfy the respective non-trivial dependence balance constraints.
\begin{figure}[t]
  \centerline{\epsfig{figure=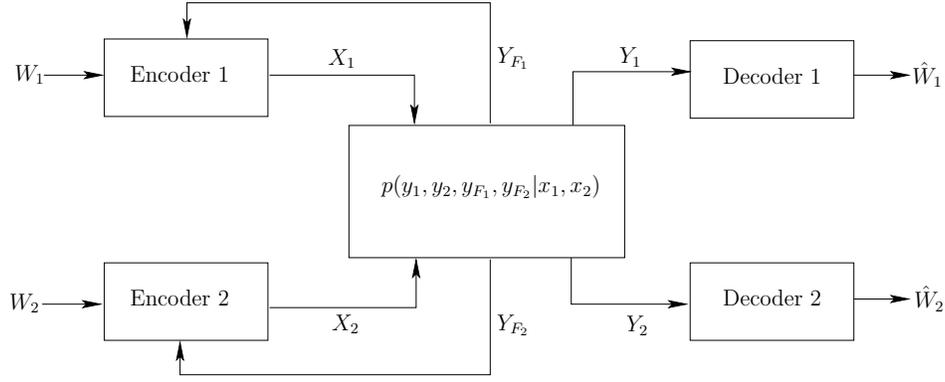,width=12.5cm}}
  \caption{The interference channel with generalized feedback (IC-GF).}\label{fig2}
\end{figure}

\section{A New Outer Bound for MAC-GF}
\begin{Theo}
The capacity region of MAC-GF is contained in the region
 \begin{align}
 \mathcal{DB}^{MAC}=\big\{(R_{1},R_{2}):\hspace{0.05in}&R_{1} \leq I(X_{1};Y,Y_{F_{2}}|X_{2},T_{2})\label{DBMAC1}\\
 &R_{2} \leq I(X_{2};Y,Y_{F_{1}}|X_{1},T_{1})\label{DBMAC2}\\
 &R_{1}+R_{2} \leq I(X_{1},X_{2};Y,Y_{F_{1}},Y_{F_{2}}|T_{1},T_{2})\label{DBMAC3}\\
 &R_{1}+R_{2}\leq I(X_{1},X_{2};Y)\big\}\label{DBMAC4}
 \end{align}
 where the random variables $(T_{1},T_{2},X_{1},X_{2},Y,Y_{F_{1}},Y_{F_{2}})$ have the joint
 distribution
 \begin{align}
 p(t_{1},t_{2},x_{1},x_{2},y,y_{F_{1}},y_{F_{2}})&=p(t_{1},t_{2},x_{1},x_{2})p(y,y_{F_{1}},y_{F_{2}}|x_{1},x_{2})\label{DBMAC5}
 \end{align}
 and also satisfy the following dependence balance bound
 \begin{align}
 I(X_{1};X_{2}|T_{1},T_{2})&\leq I(X_{1};X_{2}|Y_{F_{1}},Y_{F_{2}},T_{1},T_{2})\label{DBMAC6}
 \end{align}
 \end{Theo}
The proof of Theorem $1$ is given in the Appendix.
\section{A New Outer Bound for IC-GF}

\begin{Theo}
The capacity region of IC-GF is contained in the region
\begin{align}
\mathcal{DB}^{IC}=\big\{(R_{1},R_{2}):\hspace{0.05in}&R_{1}\leq I(X_{1},X_{2};Y_{1})\label{DBIC1}\\
&R_{2}\leq I(X_{1},X_{2};Y_{2})\label{DBIC2}\\
&R_{1}\leq I(X_{1};Y_{1},Y_{2},Y_{F_{2}}|X_{2},T_{2})\label{DBIC3}\\
&R_{2}\leq I(X_{2};Y_{1},Y_{2},Y_{F_{1}}|X_{1},T_{1})\label{DBIC4}\\
&R_{1}+R_{2}\leq I(X_{1},X_{2};Y_{1},Y_{2},Y_{F_{1}},Y_{F_{2}}|T_{1},T_{2})\label{DBIC5}\\
&R_{1}+R_{2}\leq I(X_{1},X_{2};Y_{1},Y_{2})\big\}\label{DBIC6}
\end{align}
where the random variables $(T_{1},T_{2},X_{1},X_{2},Y_{1},Y_{2},Y_{F_{1}},Y_{F_{2}})$ have the joint
distribution
\begin{align}
p(t_{1},t_{2},x_{1},x_{2},y_{1},y_{2},y_{F_{1}},y_{F_{2}})&=p(t_{1},t_{2},x_{1},x_{2})p(y_{1},y_{2},y_{F_{1}},y_{F_{2}}|x_{1},x_{2})\label{DBIC7}
\end{align}
and also satisfy the following dependence balance bound
\begin{align}
I(X_{1};X_{2}|T_{1},T_{2})&\leq I(X_{1};X_{2}|Y_{F_{1}},Y_{F_{2}},T_{1},T_{2})\label{DBIC8}
\end{align}
\end{Theo}
The proof of Theorem $2$ is given in the Appendix.

We note here that one can obtain fixed and adaptive parallel
channel extensions of the dependence balance based bounds in a similar
fashion as in \cite{Hekstra_Willems:1989}. The parallel channel
extensions could potentially improve upon the outer bounds derived
in this paper. For the scope of this paper, we will only use
Theorems $1$ and $2$. In the next three sections, we will consider specific
channel models of MAC with noisy feedback, MAC with user
cooperation, and IC with user cooperation and specialize Theorems $1$ and $2$
for these channel models.
In particular, we will show that for these three channel models, it
is sufficient to employ a single auxiliary random variable $T$, as
opposed to two auxiliary random variables $T_{1}$ and $T_{2}$
appearing in Theorems $1$ and $2$.

We should also remark here that dependence balance approach was
first applied by Gastpar and Kramer for the Gaussian MAC with
noisy feedback in \cite{GK1:2006} and for the Gaussian IC with
noisy feedback (IC-NF) in \cite{GK3:2006}. An interesting
Lagrangian based approach was proposed in \cite{GK3:2006} to
partially evaluate the dependence balance based outer bound for
the Gaussian IC-NF and it was shown that dependence balance based
bounds strictly improve upon the cut-set outer bound. For this
reason, we do not consider the Gaussian IC-NF in this paper.

\section{Gaussian MAC with Noisy Feedback}
We first consider the Gaussian MAC with noisy feedback (see Figure $3$).
The channel model is given as,
\begin{align}
Y&=X_{1}+X_{2}+Z\label{NF1}\\
Y_{F_{1}}&=Y+Z_{1}\label{NF2}\\
Y_{F_{2}}&=Y+Z_{2}\label{NF3}
\end{align}
where $Z$,$Z_{1}$ and $Z_{2}$ are independent, zero-mean, Gaussian
random variables with variances $\sigma_{Z}^{2}$,
$\sigma_{Z_{1}}^{2}$ and $\sigma_{Z_{2}}^{2}$, respectively.
Moreover, the channel inputs are subject to average power
constraints, $E[X_{1}^{2}]\leq P_{1}$ and $E[X_{2}^{2}]\leq
P_{2}$. Note that the channel model described above has a special
probability structure, namely,
\begin{align}
p(y,y_{F_{1}},y_{F_{2}}|x_{1},x_{2})&=p(y|x_{1},x_{2})p(y_{F_{1}}|y)p(y_{F_{2}}|y)\label{channelstructureNF}
\end{align}
For any MAC-GF with a transition probability in the form of
(\ref{channelstructureNF}), we have the following strengthened
version of Theorem $1$.

\begin{Theo}
The capacity region of any MAC-GF, with a transition probability
in the form of (\ref{channelstructureNF}), is contained in the
region
 \begin{align}
 \mathcal{DB}_{NF}^{MAC}=\big\{(R_{1},R_{2}):\hspace{0.05in}&R_{1} \leq I(X_{1};Y|X_{2},T)\label{DBNF1}\\
 &R_{2} \leq I(X_{2};Y|X_{1},T)\label{DBNF2}\\
 &R_{1}+R_{2} \leq I(X_{1},X_{2};Y|T)\label{DBNF3}\big\}
 \end{align}
 where the random variables $(T,X_{1},X_{2},Y,Y_{F_{1}},Y_{F_{2}})$ have the joint
 distribution
 \begin{align}
 p(t,x_{1},x_{2},y,y_{F_{1}},y_{F_{2}})&=p(t,x_{1},x_{2})p(y|x_{1},x_{2})p(y_{F_{1}}|y)p(y_{F_{2}}|y)\label{DBNF4}
 \end{align}
 and  also satisfy the following dependence balance bound
 \begin{align}
 I(X_{1};X_{2}|T)&\leq I(X_{1};X_{2}|Y_{F_{1}},Y_{F_{2}},T)\label{DBNF5}
 \end{align}
where the random variable $T$ is subject to a cardinality constraint
$|\mathcal{T}|\leq |\mathcal{X}_{1}||\mathcal{X}_{2}|+3$.
\end{Theo}
The proof of Theorem $3$ is given in the Appendix.

In Section $10$, we will show that it suffices to consider jointly
Gaussian $(T,X_{1},X_{2})$ satisfying (\ref{DBNF5}) when evaluating
Theorem $3$ for the Gaussian MAC with noisy feedback described in
(\ref{NF1})-(\ref{NF3}).

\begin{figure}[t]
  \centerline{\epsfig{figure=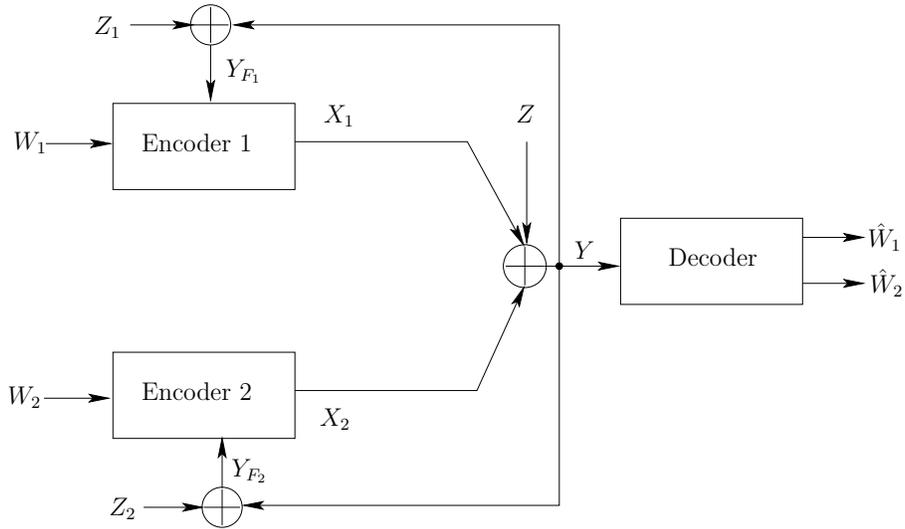,width=12.0cm}}
   \caption{The Gaussian MAC with noisy feedback.}\label{fig3}
\end{figure}

\section{Gaussian MAC with User Cooperation}
In this section, we consider the Gaussian MAC with user cooperation
\cite{Aazhang:2003}, where each transmitter receives a noisy version of the other
transmitter's channel input. The user cooperation model (see Figure $4$) is a special instance of a MAC-GF, where
the channel outputs are described as,
\begin{align}
Y&=\sqrt{h_{10}}X_{1}+\sqrt{h_{20}}X_{2}+Z\label{UCmodel1}\\
Y_{F_{1}}&=\sqrt{h_{21}}X_{2}+Z_{1}\label{UCmodel2}\\
Y_{F_{2}}&=\sqrt{h_{12}}X_{1}+Z_{2}\label{UCmodel3}
\end{align}
where $Z$,$Z_{1}$ and $Z_{2}$ are independent, zero-mean, Gaussian
random variables with variances $\sigma_{Z}^{2}$,
$\sigma_{Z_{1}}^{2}$ and $\sigma_{Z_{2}}^{2}$, respectively. The
channel gains $h_{10},h_{20}, h_{12}$ and $h_{21}$ are assumed to
be fixed and known at all terminals. Moreover, the channel inputs
are subject to average power constraints, $E[X_{1}^{2}]\leq P_{1}$
and $E[X_{2}^{2}]\leq P_{2}$. Note that the channel model
described above has a special probability structure, namely,
\begin{align}
p(y,y_{F_{1}},y_{F_{2}}|x_{1},x_{2})&=p(y|x_{1},x_{2})p(y_{F_{1}}|x_{2})p(y_{F_{2}}|x_{1})\label{channelstructure}
\end{align}
For any MAC-GF with a transition probability in the form of
(\ref{channelstructure}), we have the following strengthened
version of Theorem $1$.

\begin{Theo}
The capacity region of any MAC-GF with a transition probability in
the form of (\ref{channelstructure}), is contained in the region
 \begin{align}
 \mathcal{DB}_{UC}^{MAC}=\big\{(R_{1},R_{2}):\hspace{0.05in}&R_{1} \leq I(X_{1};Y,Y_{F_{2}}|X_{2},T)\label{DBT1}\\
 &R_{2} \leq I(X_{2};Y,Y_{F_{1}}|X_{1},T)\label{DBT2}\\
 &R_{1}+R_{2} \leq I(X_{1},X_{2};Y,Y_{F_{1}},Y_{F_{2}}|T)\label{DBT3}\\
 &R_{1}+R_{2}\leq I(X_{1},X_{2};Y)\big\}\label{DBT4}
 \end{align}
 where the random variables $(T,X_{1},X_{2},Y,Y_{F_{1}},Y_{F_{2}})$ have the joint
 distribution
 \begin{align}
 p(t,x_{1},x_{2},y,y_{F_{1}},y_{F_{2}})&=p(t,x_{1},x_{2})p(y|x_{1},x_{2})p(y_{F_{1}}|x_{2})p(y_{F_{2}}|x_{1})\label{DBT5}
 \end{align}
 and also satisfy the following dependence balance bound
 \begin{align}
 I(X_{1};X_{2}|T)&\leq I(X_{1};X_{2}|Y_{F_{1}},Y_{F_{2}},T)\label{DBT6}
 \end{align}
where the random variable $T$ is subject to a cardinality constraint
$|\mathcal{T}|\leq |\mathcal{X}_{1}||\mathcal{X}_{2}|+3$.
\end{Theo}
The proof of Theorem $4$ is given in the Appendix.

In Section $11$, we will show that it suffices to consider jointly
Gaussian $(T,X_{1},X_{2})$ satisfying (\ref{DBT6}) when evaluating
Theorem $4$ for the Gaussian MAC with user cooperation described in (\ref{UCmodel1})-(\ref{UCmodel3}).
\begin{figure}[t]
  \centerline{\epsfig{figure=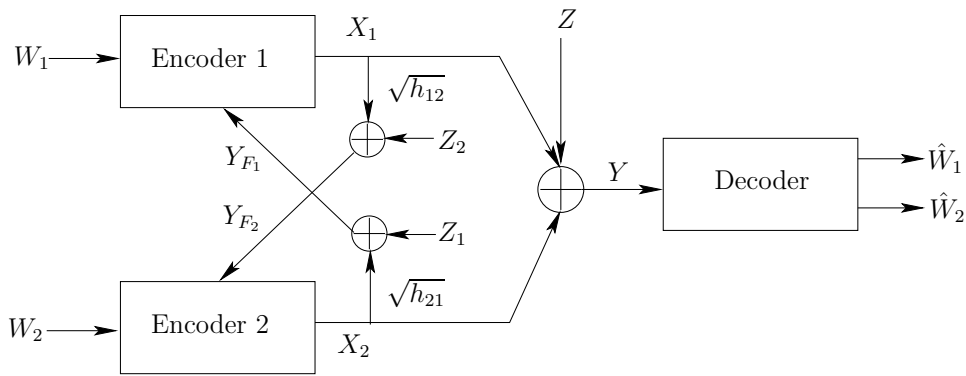,width=12.8cm}}
   \caption{The Gaussian MAC with user cooperation.}\label{fig4}
\end{figure}

\section{Gaussian IC with User Cooperation}
In this section, we will evaluate our outer bound for a user cooperation
setting \cite{Tuninetti:2007},\cite{Host-Madsen:2006}, where the transmitters receive noisy versions of the other
transmitter's channel input. The user cooperation model (see Figure $5$) is a special instance of an IC-GF, where
the channel outputs are described as,
\begin{align}
Y_{1}&=X_{1}+\sqrt{b}X_{2}+N_{1}\label{ICUCmodel1}\\
Y_{2}&=\sqrt{a}X_{1}+X_{2}+N_{2}\label{ICUCmodel2}\\
Y_{F_{1}}&=\sqrt{h_{21}}X_{2}+Z_{1}\label{ICUCmodel3}\\
Y_{F_{2}}&=\sqrt{h_{12}}X_{1}+Z_{2}\label{ICUCmodel4}
\end{align}
where $N_{1},N_{2}$, $Z_{1}$ and $Z_{2}$ are independent, zero-mean, Gaussian random
variables with variances $\sigma_{N_{1}}^{2},\sigma_{N_{2}}^{2}$, $\sigma_{Z_{1}}^{2}$ and
$\sigma_{Z_{2}}^{2}$, respectively. The channel gains $a,b, h_{12}$ and $h_{21}$
are assumed to be fixed and known at all terminals.
Moreover, the channel inputs are
subject to average power constraints, $E[X_{1}^{2}]\leq P_{1}$ and  $E[X_{2}^{2}]\leq P_{2}$.
Note that the channel model described above has a special probability
structure, namely,
\begin{align}
p(y_{1},y_{2},y_{F_{1}},y_{F_{2}}|x_{1},x_{2})&=p(y_{1},y_{2}|x_{1},x_{2})p(y_{F_{1}}|x_{2})p(y_{F_{2}}|x_{1})\label{channelstructureIC}
\end{align}
For any IC-GF with a transition probability in the form
of (\ref{channelstructureIC}), we have the following strengthened
version of Theorem $2$.

\begin{Theo}
The capacity region of any IC-GF with a transition probability in the form of (\ref{channelstructureIC}), is contained in the region
\begin{align}
\mathcal{DB}_{UC}^{IC}=\big\{(R_{1},R_{2}):\hspace{0.05in}&R_{1}\leq I(X_{1},X_{2};Y_{1})\label{DBICT1}\\
&R_{2}\leq I(X_{1},X_{2};Y_{2})\label{DBICT2}\\
&R_{1}\leq I(X_{1};Y_{1},Y_{2},Y_{F_{2}}|X_{2},T)\label{DBICT3}\\
&R_{2}\leq I(X_{2};Y_{1},Y_{2},Y_{F_{1}}|X_{1},T)\label{DBICT4}\\
&R_{1}+R_{2}\leq I(X_{1},X_{2};Y_{1},Y_{2},Y_{F_{1}},Y_{F_{2}}|T)\label{DBICT5}\\
&R_{1}+R_{2}\leq I(X_{1},X_{2};Y_{1},Y_{2})\big\}\label{DBICT6}
 \end{align}
where the random variables $(T,X_{1},X_{2},Y_{1},Y_{2},Y_{F_{1}},Y_{F_{2}})$ have the joint
distribution
\begin{align}
p(t,x_{1},x_{2},y_{1},y_{2},y_{F_{1}},y_{F_{2}})&=p(t,x_{1},x_{2})p(y_{1},y_{2}|x_{1},x_{2})p(y_{F_{1}}|x_{2})p(y_{F_{2}}|x_{1})\label{DBICT7}
\end{align}
and also satisfy the following dependence balance bound
\begin{align}
I(X_{1};X_{2}|T)&\leq I(X_{1};X_{2}|Y_{F_{1}},Y_{F_{2}},T)\label{DBICT8}
\end{align}
where the random variable $T$ is subject to a cardinality constraint
$|\mathcal{T}|\leq |\mathcal{X}_{1}||\mathcal{X}_{2}|+3$.
\end{Theo}
The proof of Theorem $5$ is given in the Appendix.

In Section $12$, we will show that it suffices to consider jointly
Gaussian $(T,X_{1},X_{2})$ satisfying (\ref{DBICT8}) when evaluating
Theorem $5$ for the Gaussian IC with user cooperation described in (\ref{ICUCmodel1})-(\ref{ICUCmodel4}).

\begin{figure}[t]
  \centerline{\epsfig{figure=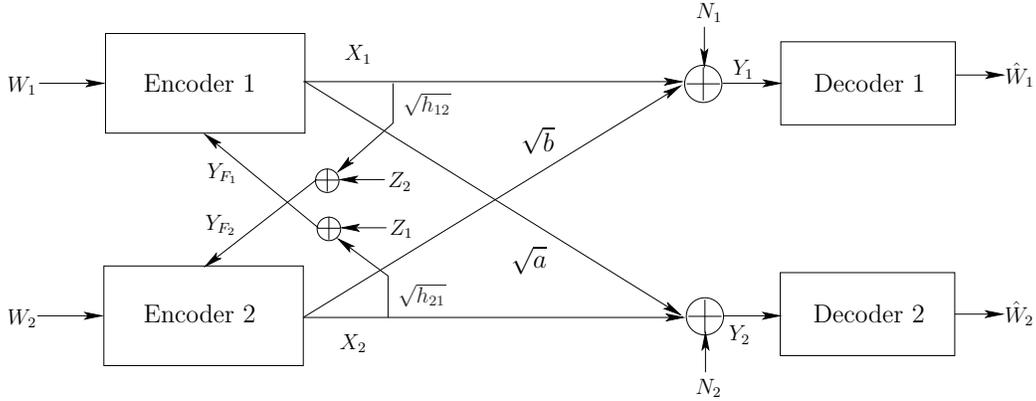,width=13.7cm}}
   \caption{The Gaussian IC with user cooperation.}\label{fig5}
\end{figure}

\section{Outline for Evaluating $\mathcal{DB}_{NF}^{MAC}$, $\mathcal{DB}_{UC}^{MAC}$ and $\mathcal{DB}_{UC}^{IC}$}

In this section, we outline the common approach for evaluation of
our outer bounds, $\mathcal{DB}_{NF}^{MAC}$ for the Gaussian MAC with
noisy feedback, $\mathcal{DB}_{UC}^{MAC}$ for the Gaussian MAC with
user-cooperation and $\mathcal{DB}_{UC}^{IC}$ for the Gaussian IC with
user-cooperation. The main difficulty in evaluating these bounds
is to identify the optimal selection of joint densities of
$(T,X_{1},X_{2})$. Our aim will be to prove that it is sufficient
to consider jointly Gaussian $(T,X_{1},X_{2})$ satisfying
(\ref{DBNF5}) for MAC with noisy feedback, (\ref{DBT6}) for MAC with user
cooperation, and (\ref{DBICT8}) for IC with user
cooperation, respectively, when evaluating the corresponding outer
bounds.

First note that the three outer bounds, namely $\mathcal{DB}_{NF}^{MAC}$, $\mathcal{DB}_{UC}^{MAC}$  and
$\mathcal{DB}_{UC}^{IC}$ have a similar structure, i.e., all outer
bounds involve taking a union over joint densities of
$(T,X_{1},X_{2})$ satisfying the constraints (\ref{DBNF5}),
(\ref{DBT6}) and (\ref{DBICT8}), respectively. Let us symbolically denote these
constraints as a variable $(DB)$, where $(DB)=$ (\ref{DBNF5}) for
MAC with noisy feedback, $(DB)=$ (\ref{DBT6}) for MAC with user cooperation,
and $(DB)=$ (\ref{DBICT8}) for IC with user cooperation.

We begin by considering the set of all distributions of three
random variables $(T,X_{1},X_{2})$ which satisfy the power
constraints, $E\big[X_{1}^{2}\big]\leq P_{1}$ and
$E\big[X_{2}^{2}\big]\leq P_{2}$. Let us formally define this set
of input distributions as
\begin{align}
\mathcal{P}&=\{p(t,x_{1},x_{2}): E\big[X_{1}^{2}\big]\leq P_{1},
E\big[X_{2}^{2}\big]\leq P_{2}\}\nonumber
\end{align}
For simplicity, we abbreviate jointly Gaussian distributions as
$\mathcal{JG}$ and distributions which are not jointly Gaussian as
$\mathcal{NG}$. We first partition $\mathcal{P}$ into two disjoint
subsets,
\begin{align}
\mathcal{P}_{G}&=\{p(t,x_{1},x_{2}) \in \mathcal{P}: (T,X_{1},X_{2}) \mbox{ are $\mathcal{JG}$}\}\nonumber\\
 \mathcal{P}_{NG}&=\{p(t,x_{1},x_{2}) \in \mathcal{P} : (T,X_{1},X_{2})
\mbox{ are $\mathcal{NG}$}\}\nonumber
\end{align}
We further individually partition the sets $\mathcal{P}_{G}$ and
$\mathcal{P}_{NG}$, respectively, as
\begin{align}
\mathcal{P}_{G}^{DB}&=\{p(t,x_{1},x_{2}) \in \mathcal{P}_{G} : (T,X_{1},X_{2}) \mbox{ satisfy $(DB)$}\}\nonumber\\
\mathcal{P}_{G}^{\overline{DB}}&=\{p(t,x_{1},x_{2}) \in
\mathcal{P}_{G} : (T,X_{1},X_{2}) \mbox{ do not satisfy
$(DB)$}\}\nonumber
\end{align}
and
\begin{align}
\mathcal{P}_{NG}^{DB}&=\{p(t,x_{1},x_{2}) \in \mathcal{P}_{NG} : (T,X_{1},X_{2}) \mbox{ satisfy $(DB)$}\}\nonumber\\
\mathcal{P}_{NG}^{\overline{DB}}&=\{p(t,x_{1},x_{2}) \in
\mathcal{P}_{NG} : (T,X_{1},X_{2}) \mbox{ do not satisfy
$(DB)$}\}\nonumber
\end{align}
Finally, we partition the set $\mathcal{P}_{NG}^{DB}$ into two
disjoint sets $\mathcal{P}_{NG}^{DB(a)}$ and
$\mathcal{P}_{NG}^{DB(b)}$ with
$\mathcal{P}_{NG}^{DB}=\mathcal{P}_{NG}^{DB(a)}\bigcup\mathcal{P}_{NG}^{DB(b)}$,
as
\begin{align}
\mathcal{P}_{NG}^{DB(a)}=\big\{&p(t,x_{1},x_{2})\in
\mathcal{P}_{NG}^{DB}: \mbox{ covariance matrix of } p(t,x_{1},x_{2}) \mbox{ is } Q \mbox{ and there } \nonumber\\
&  \mbox{exists a $\mathcal{JG}$ } (T_{G},X_{1G},X_{2G}) \mbox{ with
  covariance matrix } Q  \mbox{ satisfying } (DB)\big\}\nonumber\\
\mathcal{P}_{NG}^{DB(b)}=\big\{&p(t,x_{1},x_{2})\in
\mathcal{P}_{NG}^{DB}: \mbox{ covariance matrix of }
p(t,x_{1},x_{2}) \mbox{ is } Q \mbox{ and there } \nonumber\\&
\mbox{does not exist a $\mathcal{JG}$ } (T_{G},X_{1G},X_{2G})
\mbox{
  with covariance matrix } Q\mbox{ satisfying }(DB)\big\}\nonumber
\end{align}

So far, we have partitioned the set of input distributions into
five disjoint sets: $\mathcal{P}_{G}^{DB}$,
$\mathcal{P}_{G}^{\overline{DB}}$, $\mathcal{P}_{NG}^{DB(a)}$,
$\mathcal{P}_{NG}^{DB(b)}$ and $\mathcal{P}_{NG}^{\overline{DB}}$.
To visualize this partition of the set of input distributions, see
Figure $6$. It is clear that the input distributions which fall
into the sets $\mathcal{P}_{G}^{\overline{DB}}$ and
$\mathcal{P}_{NG}^{\overline{DB}}$ need not be considered since
they do not satisfy the constraint $(DB)$ and do not have any
consequence when evaluating our outer bounds. Therefore, we only
need to restrict our attention on the three remaining sets
$\mathcal{P}_{G}^{DB}$, $\mathcal{P}_{NG}^{DB(a)}$, and
$\mathcal{P}_{NG}^{DB(b)}$ i.e., those input distributions which
satisfy the dependence balance bound.

We explicitly evaluate our outer bound in the following three
steps:
\begin{enumerate}
\item We first explicitly characterize the region of rate pairs
provided by our outer bound for the probability distributions in
the set $\mathcal{P}_{G}^{DB}$.

\item In the second step, we will show that for any input
distribution belonging to the set $\mathcal{P}_{NG}^{DB(a)}$,
there exists an input distribution in the set
$\mathcal{P}_{G}^{DB}$ which yields a set of larger rate pairs.
This leads to the conclusion that we do not need to consider the
input distributions in the set $\mathcal{P}_{NG}^{DB(a)}$ in
evaluating our outer bound.

\item We next focus on the set $\mathcal{P}_{NG}^{DB(b)}$ and show
that for any non-Gaussian input distribution $p(t,x_{1},x_{2})\in
\mathcal{P}_{NG}^{DB(b)}$, we can construct a jointly Gaussian
input distribution satisfying $(DB)$, i.e., we can find a
corresponding input distribution in $\mathcal{P}_{G}^{DB}$, which
yields a set of rates which includes the set of
rates of the fixed non-Gaussian input distribution
$p(t,x_{1},x_{2})$.
\end{enumerate}

The main step in evaluating our outer bounds is step $3$ described
above. The proofs of step $3$ for noisy feedback and user
cooperation models are entirely different and do not follow from
each other. The evaluation in step $1$ is slightly different for
all three settings, also owing to the channel models. Hence, we will
separately focus on these models in the following three sections.
\begin{figure}[t]
  \centerline{\epsfig{figure=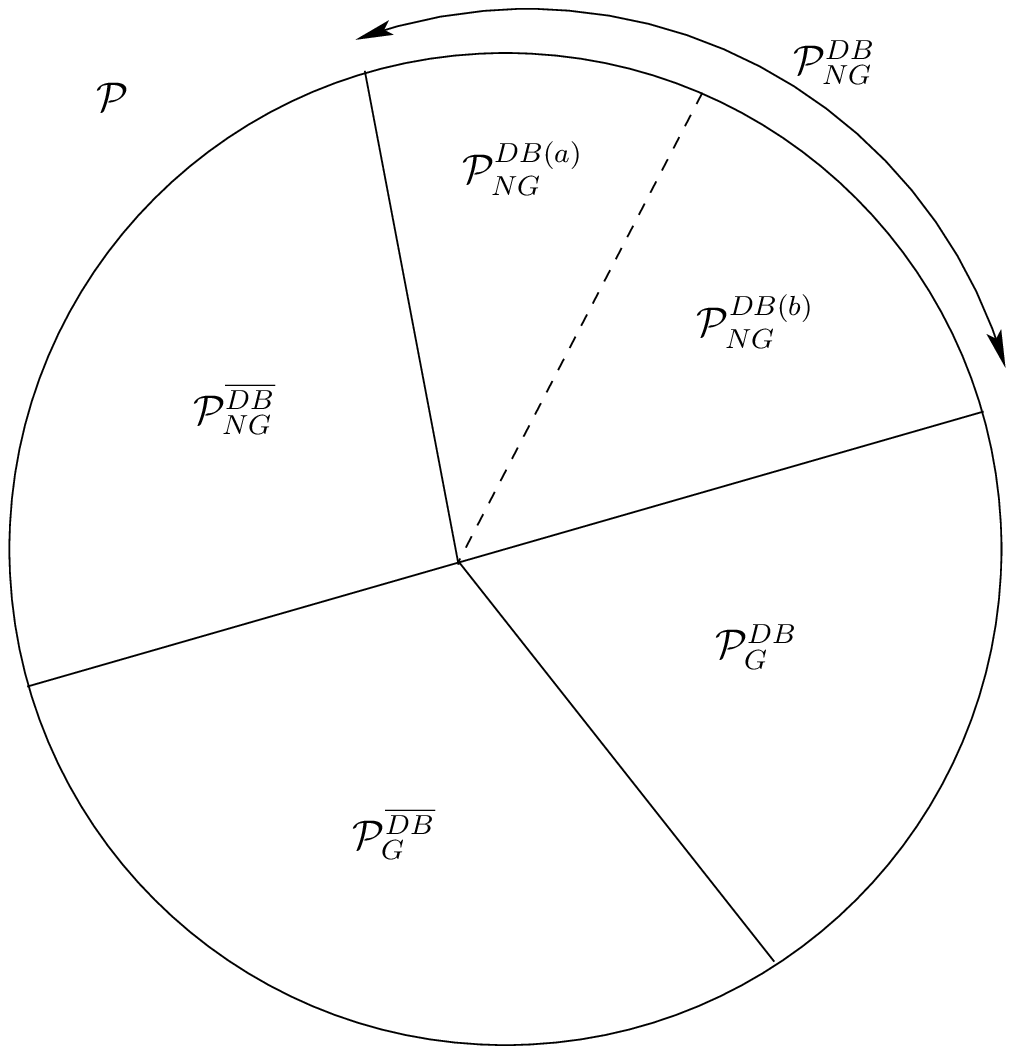,width=7.5cm}}
  \caption{A partition of the set of input distributions $\mathcal{P}$.}\label{fig6}
\end{figure}
Contrary to steps $1$ and $3$, step $2$ is common for all channel
models. Therefore, we first present the common result for all
channel models here. In step $2$, we consider any non-Gaussian
input distribution $p(t,x_{1},x_{2})$ in
$\mathcal{P}_{NG}^{DB(a)}$ with a covariance matrix $Q$. For such
an input distribution, we know by the maximum entropy theorem
\cite{Cover:book}, that the rates provided by a jointly Gaussian
triple with the same covariance matrix $Q$ are always at least as
large as the rates provided by the chosen non-Gaussian
distribution. Therefore, for any input distribution in
$\mathcal{P}_{NG}^{DB(a)}$, there always exists an input
distribution in $\mathcal{P}_{G}^{DB}$, satisfying $(DB)$, which
yields larger rates. This means that we can ignore the set
$\mathcal{P}_{NG}^{DB(a)}$ altogether while evaluating our outer
bounds.

To set the stage for our evaluations in steps $1$ and $3$ for the
three channel models, let us define $\mathcal{Q}$ as the set of all
valid $3\times 3$ covariance matrices of three random variables
$(T,X_{1},X_{2})$. A typical element $Q$ in the set $\mathcal{Q}$
takes the following form,
\begin{align}
Q&=E\big[(X_{1}\hspace{0.03in} X_{2}
\hspace{0.03in}T)(X_{1}\hspace{0.03in} X_{2}
\hspace{0.03in}T)^{T}\big]\nonumber\\
&=\left(%
\begin{array}{ccc}
  P_{1} & \rho_{12}\sqrt{P_{1}P_{2}} & \rho_{1T}\sqrt{P_{1}P_{T}} \\
  \rho_{12}\sqrt{P_{1}P_{2}} & P_{2} & \rho_{2T}\sqrt{P_{2}P_{T}} \\
  \rho_{1T}\sqrt{P_{1}P_{T}} & \rho_{2T}\sqrt{P_{2}P_{T}} & P_{T} \\
\end{array}%
\right)\label{Qdef1}
\end{align}
A necessary condition for $Q$ to be a valid covariance matrix is
that it is positive semi-definite, i.e., $\mbox{det}(Q)\geq 0$.
This is equivalent to saying that,
\begin{align}
\mbox{det}(Q)
&=P_{1}P_{2}P_{T}\Delta \geq 0\label{Qdef2}
\end{align}
where we have defined for simplicity,
\begin{align}
\Delta&=1-\rho_{12}^{2}-\rho_{1T}^{2}-\rho_{2T}^{2}+2\rho_{1T}\rho_{2T}\rho_{12}\label{Qdef3}
\end{align}

\section{Evaluation of $\mathcal{DB}_{NF}^{MAC}$}

In this section we explicitly evaluate Theorem $3$ for the
Gaussian MAC with noisy feedback described by
(\ref{NF1})-(\ref{NF3}) in Section $6$. We start with step $1$. We consider an
input distribution in $\mathcal{P}_{G}^{DB}$, i.e., a jointly
Gaussian triple $(T_{G},X_{1G},X_{2G})$ with a covariance matrix
$Q$. Let us first characterize the set of rate constraints for
this triple. It is straightforward to evaluate the three rate
constraints appearing in (\ref{DBNF1})-(\ref{DBNF3}) for this
input distribution
\begin{align}
R_{1}&\leq I(X_{1G};Y|X_{2G},T_{G})=\frac{1}{2}\mbox{log}\left(1+\frac{f_{1}(Q)}{\sigma_{Z}^{2}}\right)\label{E13}\\
R_{2}&\leq I(X_{2G};Y|X_{1G},T_{G})=\frac{1}{2}\mbox{log}\left(1+\frac{f_{2}(Q)}{\sigma_{Z}^{2}}\right)\label{E14}\\
R_{1}+R_{2}&\leq
I(X_{1G},X_{2G};Y|T_{G})=\frac{1}{2}\mbox{log}\left(1+\frac{f_{3}(Q)}{\sigma_{Z}^{2}}\right)\label{E15}
\end{align}
where we have defined
\begin{align}
f_{1}(Q)&=\mbox{Var}(X_{1G}|X_{2G},T_{G})=\frac{\Delta P_{1}}{(1-\rho_{2T}^{2})}\label{E16}\\
f_{2}(Q)&=\mbox{Var}(X_{2G}|X_{1G},T_{G})=\frac{\Delta
P_{2}}{(1-\rho_{1T}^{2})}\label{E17}\\
f_{3}(Q)&=\mbox{Var}(X_{1G}|T_{G})+\mbox{Var}(X_{2G}|T_{G})+2\mbox{Cov}(X_{1G},X_{2G}|T_{G})\nonumber\\
&=(1-\rho_{1T}^{2})P_{1}+(1-\rho_{2T}^{2})P_{2}+2(\rho_{12}-\rho_{1T}\rho_{2T})\sqrt{P_{1}P_{2}}\label{E18}
\end{align}
Finally, evaluating the constraint in (\ref{DBNF5}), we conclude
that this input distribution satisfies the constraint in
(\ref{DBNF5}) iff,
\begin{align}
f_{3}(Q)&\leq f_{1}(Q)+f_{2}(Q) +
\frac{f_{1}(Q)f_{2}(Q)}{\left(\sigma_{Z}^{2}+\frac{\sigma_{Z_{1}}^{2}\sigma_{Z_{2}}^{2}}{(\sigma_{Z_{1}}^{2}+\sigma_{Z_{2}}^{2})}\right)}\label{E22}
\end{align}
To summarize, the set of rate pairs provided by an input
distribution in $\mathcal{P}_{G}^{DB}$, with a covariance matrix
$Q$, are given by those in (\ref{E13})-(\ref{E15}), where
$f_{i}(Q)$, $i=1,2,3$, in those inequalities are subject to the
constraint in (\ref{E22}). As we have discussed earlier, from
evaluation of step $2$ in Section $9$, we know that all rate pairs
contributed by input distributions in $\mathcal{P}_{NG}^{{DB(a)}}$
are covered by those given in $\mathcal{P}_{G}^{DB}$.

We now arrive at step $3$ of our evaluation. Consider any input
distribution $p(t,x_{1},x_{2})$ in $\mathcal{P}_{NG}^{DB(b)}$ with
a covariance matrix $Q$. By the definition of the set
$\mathcal{P}_{NG}^{DB(b)}$, we know that $Q$ does not satisfy
(\ref{DBNF5}), which implies
\begin{align}
f_{3}(Q)> f_{1}(Q)+f_{2}(Q) +
\frac{f_{1}(Q)f_{2}(Q)}{\left(\sigma_{Z}^{2}+\frac{\sigma_{Z_{1}}^{2}\sigma_{Z_{2}}^{2}}{(\sigma_{Z_{1}}^{2}+\sigma_{Z_{2}}^{2})}\right)}\label{E23}
\end{align}
We also note that for any $(T,X_{1},X_{2})$ with a covariance
matrix $Q$,
\begin{align}
R_{1}&\leq I(X_{1};Y|X_{2},T)\leq
\frac{1}{2}\mbox{log}\left(1+\frac{f_{1}(Q)}{\sigma_{Z}^{2}}\right)\label{E24}\\
R_{2}&\leq I(X_{2};Y|X_{1},T)\leq
\frac{1}{2}\mbox{log}\left(1+\frac{f_{2}(Q)}{\sigma_{Z}^{2}}\right)\label{E25}\\
R_{1}+R_{2}&\leq I(X_{1},X_{2};Y|T)\leq
\frac{1}{2}\mbox{log}\left(1+\frac{f_{3}(Q)}{\sigma_{Z}^{2}}\right)\label{E26}
\end{align}
which is a simple consequence of the maximum entropy theorem
\cite{Cover:book}. Note that so far, we have not used the fact
that the given non-Gaussian input distribution satisfies the
dependence balance constraint in (\ref{DBNF5}). We will now make
use of this fact by rewriting (\ref{DBNF5}) as follows,
\begin{align}
0&\leq I(X_{1};X_{2}|Y_{F_{1}},Y_{F_{2}},T)-I(X_{1};X_{2}|T)\label{E27}\\
&= I(X_{1};Y_{F_{1}},Y_{F_{2}}|X_{2},T)-I(X_{1};Y_{F_{1}},Y_{F_{2}}|T)\label{E28}\\
&=
 h(Y_{F_{1}},Y_{F_{2}}|X_{1},T)+h(Y_{F_{1}},Y_{F_{2}}|X_{2},T)-h(Y_{F_{1}},Y_{F_{2}}|T)-h(Y_{F_{1}},Y_{F_{2}}|X_{1},X_{2},T)\label{E29}
\end{align}
We express the above constraint as,
\begin{align}
h(Y_{F_{1}},Y_{F_{2}}|T)+h(Y_{F_{1}},Y_{F_{2}}|X_{1},X_{2},T)&\leq
h(Y_{F_{1}},Y_{F_{2}}|X_{1},T)+h(Y_{F_{1}},Y_{F_{2}}|X_{2},T)\label{E30}
\end{align}
Before proceeding, we state a recently discovered multivariate
generalization \cite{Palomar:2008} of Costa's EPI \cite{Costa:EPI1985}.

\begin{Lem}
For any arbitrary random vector $\mathbf{Y} \in \mathbb{R}^{2}$,
independent of $\mathbf{V}\in \mathbb{R}^{2}$, where $\mathbf{V}$
is a zero-mean, Gaussian random vector with each component having
unit variance, the entropy power
$N(\Lambda^{1/2}\mathbf{Y}+\mathbf{V})$ is concave in $\Lambda$,
where the entropy power is defined as
\begin{align}
N(\mathbf{Y})&=\frac{1}{(2\pi\mbox{e})}\mbox{e}^{h(\mathbf{Y})}\label{defEP}
\end{align}
and $\Lambda$ is a diagonal matrix with components
$(\lambda_{1},\lambda_{2})$.
\end{Lem}

We can therefore write for any pair of diagonal matrices
$\Lambda_{1}, \Lambda_{2}$ and for any $\mu\in [0,1]$,
\begin{align}
\mu N(\Lambda_{1}^{1/2}\mathbf{Y}+\mathbf{V})+(1-\mu)
N(\Lambda_{2}^{1/2}\mathbf{Y}+\mathbf{V})&\leq
N((\mu\Lambda_{1}+(1-\mu)\Lambda_{2})^{1/2}\mathbf{Y}+\mathbf{V})\label{Palomar}
\end{align}
We start by obtaining a lower bound for the first term
$h(Y_{F_{1}},Y_{F_{2}}|T)$ in (\ref{E30}),
\begin{align}
h(Y_{F_{1}},Y_{F_{2}}|T)
&=\int f_{T}(t)h(Y+Z_{1},Y+Z_{2}|T=t)dt\label{E31}\\
&\geq \int f_{T}(t)\frac{1}{2}\mbox{log}\left((2\pi \mbox{e})^{2}\sigma_{Z_{1}}^{2}\sigma_{Z_{2}}^{2}+2\pi\mbox{e}(\sigma_{Z_{1}}^{2}+\sigma_{Z_{2}}^{2})\mbox{e}^{2h(Y|T=t)}\right)dt \label{E32}\\
&\geq \frac{1}{2}\mbox{log}\left((2\pi
\mbox{e})^{2}\sigma_{Z_{1}}^{2}\sigma_{Z_{2}}^{2}+2\pi\mbox{e}(\sigma_{Z_{1}}^{2}+\sigma_{Z_{2}}^{2})\mbox{e}^{2h(Y|T)}\right)
\label{E33}
\end{align}
where (\ref{E32}) follows from the conditional version of Lemma
$1$, by selecting the following $\Lambda_{1}$, $\Lambda_{2}$ and
$\mu$
\begin{align}
\Lambda_{1}&=\left(%
\begin{array}{cc}
  \kappa & 0 \\
  0 & 0 \\
\end{array}%
\right),\quad
\Lambda_{2}=\left(%
\begin{array}{cc}
  0 & 0 \\
  0 & \kappa \\
\end{array}%
\right)\label{lambdadef}
\end{align}
where
\begin{align}
\kappa&=\frac{(\sigma_{Z_{1}}^{2}+\sigma_{Z_{2}}^{2})}{\sigma_{Z_{1}}^{2}\sigma_{Z_{2}}^{2}}\label{kappadef}
\end{align}
and
\begin{align}
\mu&=\frac{\sigma_{Z_{2}}^{2}}{(\sigma_{Z_{1}}^{2}+\sigma_{Z_{2}}^{2})}\label{mudef}
\end{align}
and by making the following substitutions,
\begin{align}
V_{1}&=\frac{Z_{1}}{\sigma_{Z_{1}}}, \quad
V_{2}=\frac{Z_{2}}{\sigma_{Z_{2}}}\label{Vdef}
\end{align}
where $\mathbf{V}=[V_{1}\quad  V_{2}]^{T}$ and $\mathbf{Y}=[Y
\quad Y]^{T}$. A derivation of (\ref{E32}) is given in the
Appendix. Next, (\ref{E33}) follows from the fact that
$\mbox{log}(\mbox{e}^{x}c_{1}+c_{2})$ is convex in $x$ for
$c_{1},c_{2}\geq 0$ and a subsequent application of Jensen's
inequality \cite{Cover:book}\footnote{ We should remark here, that
an application of the regular form of vector EPI yields the
following trivial lower bound on $h(Y_{F_{1}},Y_{F_{2}}|T)$ and
therefore, the new EPI is crucial for this step.
\begin{align}
h(Y_{F_{1}},Y_{F_{2}}|T)&\geq \frac{1}{2}\mbox{log}\left((2\pi
\mbox{e})^{2}\sigma_{Z_{1}}^{2}\sigma_{Z_{2}}^{2}\right)\nonumber
\end{align}}.

We next obtain an upper bound for the right hand side of
(\ref{E30}) by using the maximum entropy theorem as,
\begin{align}
&h(Y_{F_{1}},Y_{F_{2}}|X_{1},T)+h(Y_{F_{1}},Y_{F_{2}}|X_{2},T)\nonumber\\
&\hspace{0.4in}\leq
\frac{1}{2}\mbox{log}\left((2\pi\mbox{e})^{4}(f_{1}(Q)(\sigma_{Z_{1}}^{2}+\sigma_{Z_{2}}^{2})+\eta)(f_{2}(Q)(\sigma_{Z_{1}}^{2}+\sigma_{Z_{2}}^{2})+\eta)\right)\label{E34}
\end{align}
where we have defined
\begin{align}
\eta&=\sigma_{Z_{1}}^{2}\sigma_{Z_{2}}^{2}+
\sigma_{Z}^{2}(\sigma_{Z_{1}}^{2}+\sigma_{Z_{2}}^{2})
\end{align}
 Now, using (\ref{E30}), (\ref{E33}) and
(\ref{E34}), we obtain an upper bound on $h(Y|T)$ as follows,
\begin{align}
h(Y|T)&\leq \frac{1}{2}\mbox{log}\left((2\pi
\mbox{e})(\sigma_{Z}^{2}+f(Q))\right)\label{E35}
\end{align}
where we have defined for simplicity,
\begin{align}
f(Q)&=f_{1}(Q)+f_{2}(Q)+\frac{f_{1}(Q)f_{2}(Q)}{\left(\sigma_{Z}^{2}+\frac{\sigma_{Z_{1}}^{2}\sigma_{Z_{2}}^{2}}{(\sigma_{Z_{1}}^{2}+\sigma_{Z_{2}}^{2})}\right)}\label{E36}
\end{align}
Using (\ref{E35}), we obtain an upper bound on the sum-rate
$I(X_{1},X_{2};Y|T)$ for any non-Gaussian distribution in
$\mathcal{P}_{NG}^{DB(b)}$ as,
\begin{align}
R_{1}+R_{2}\leq I(X_{1},X_{2};Y|T)\leq
\frac{1}{2}\mbox{log}\left(1+\frac{f(Q)}{\sigma_{Z}^{2}}\right)\label{E37}
\end{align}
Comparing with (\ref{E26}) and using the fact that $Q$ satisfies
(\ref{E23}), i.e., $f(Q)<f_{3}(Q)$, we have the following set of inequalities,
\begin{align}
R_{1}+R_{2}\leq I(X_{1},X_{2};Y|T)\leq
\frac{1}{2}\mbox{log}\left(1+\frac{f(Q)}{\sigma_{Z}^{2}}\right) <
\frac{1}{2}\mbox{log}\left(1+\frac{f_{3}(Q)}{\sigma_{Z}^{2}}\right)\label{E38}
\end{align}
This leads to the observation that a combined application of the
EPI and the dependence balance bound yields a
strictly smaller upper bound for $I(X_{1},X_{2};Y|T)$ for any
distribution in $\mathcal{P}_{NG}^{DB(b)}$ than the one provided
by the maximum entropy theorem. Therefore, the rate pairs
contributed by an input distribution in $\mathcal{P}_{NG}^{DB(b)}$
with a covariance matrix $Q$ are always included in the set of
rate pairs expressed by (\ref{E24}), (\ref{E25}) and (\ref{E37}),
where $f(Q)$ is defined in (\ref{E36}).

We now arrive at the final step of our evaluation where we will show
that for this input distribution in $\mathcal{P}_{NG}^{DB(b)}$, we
can always find an input distribution in $\mathcal{P}_{G}^{DB}$, with a set of rate pairs
which include the set of rate pairs expressed by (\ref{E24}), (\ref{E25}) and (\ref{E37}).
In particular, we will show the existence of a valid covariance matrix $S$ for which the
following inequalities hold true,
\begin{align}
f_{1}(Q)&\leq f_{1}(S)\label{E39}\\
f_{2}(Q)&\leq f_{2}(S)\label{E40}\\
f(Q)&\leq f_{3}(S)\label{E41}
\end{align}
and
\begin{align}
f_{3}(S)&= f_{1}(S)+f_{2}(S) +
\frac{f_{1}(S)f_{2}(S)}{\left(\sigma_{Z}^{2}+\frac{\sigma_{Z_{1}}^{2}\sigma_{Z_{2}}^{2}}{(\sigma_{Z_{1}}^{2}+\sigma_{Z_{2}}^{2})}\right)}\label{E42}
\end{align}
Inequalities in (\ref{E39})-(\ref{E41}) will guarantee that a
Gaussian input distribution with covariance matrix $S$ yields a
larger set of rate pairs than the set of rate pairs expressed by
(\ref{E24}), (\ref{E25}) and (\ref{E37}) and the equality in
(\ref{E42}) guarantees that this input distribution satisfies the
dependence balance constraint with equality, hence it is a member
of the set $\mathcal{P}_{G}^{DB}$.

Before showing the existence of such an $S$, we first characterize
the set of covariance matrices $Q$ which satisfy (\ref{E23}).
First recall that for any $Q$ to be a valid covariance matrix, we
had the condition $\mbox{det}(Q)\geq 0$ which is equivalent to
$\Delta\geq 0$, which amounts to
\begin{align}
1-\rho_{12}^{2}-\rho_{1T}^{2}-\rho_{2T}^{2}+2\rho_{1T}\rho_{2T}\rho_{12}\geq
0\label{EE}
\end{align}
In particular, it is easy to verify that for
any given fixed pair $(\rho_{1T},\rho_{2T}) \in [-1,1]\times [-1,1]$, the set
of $\rho_{12}$ which yield a valid $Q$ are such that,
\begin{align}
\rho_{1T}\rho_{2T}-\lambda \leq \rho_{12} \leq
\rho_{1T}\rho_{2T}+\lambda \label{E43}
\end{align}
where we have defined
\begin{align}
\lambda&=\sqrt{(1-\rho_{1T}^{2})(1-\rho_{2T}^{2})}\label{E44}
\end{align}
We now consider two cases which can arise for a given
covariance matrix $Q$.

Case $1$. $Q$ is such that
    $\rho_{12}=\rho_{1T}\rho_{2T}-\alpha$, for some $\alpha \in [0,
    \lambda]$: This case is rather trivial and the following simple choice
    of $S$ works,
\begin{align}
\rho_{1T}^{(S)}&=\rho_{1T},\quad \rho_{2T}^{(S)}=\rho_{2T}\label{E45}\\
\rho_{12}^{(S)}&=\rho_{1T}\rho_{2T}\label{E46}
\end{align}
Clearly, this $S$ satisfies the dependence balance bound.
Moreover, the following inequalities hold as well,
\begin{align}
f_{1}(Q)&\leq
f_{1}(S)=(1-\rho_{1T}^{2})P_{1}\label{E47}\\
f_{2}(Q)&\leq
f_{2}(S)=(1-\rho_{2T}^{2})P_{2}\label{E48}\\
f(Q)&<f_{3}(Q)\\
&=(1-\rho_{1T}^{2})P_{1}+(1-\rho_{2T}^{2})P_{2}-2\alpha\sqrt{P_{1}P_{2}}\\
&\leq (1-\rho_{1T}^{2})P_{1}+(1-\rho_{2T}^{2})P_{2}\\
&=f_{3}(S)\label{E49}
\end{align}

Case $2$. $Q$ is such that
$\rho_{12}=\rho_{1T}\rho_{2T}+\alpha_{0}$, for some $\alpha_{0} \in
(0,\lambda]$ and $Q$ satisfies (\ref{E23}): For this case, we will construct a valid covariance matrix
$S$ as follows,
\begin{align}
\rho_{1T}^{(S)}&=\rho_{1T},\quad
\rho_{2T}^{(S)}=\rho_{2T}\label{E50}\\
\rho_{12}^{(S)}&=\rho_{1T}\rho_{2T}+\alpha^{*}, \qquad \mbox{for some }
0< \alpha^{*} < \alpha_{0} \label{E51}
\end{align}
We define a parameterized covariance matrix $Q(\alpha)$ with entries,
\begin{align}
\rho_{1T}(\alpha)&=\rho_{1T},\quad   \rho_{2T}(\alpha)=\rho_{2T}\label{E53}\\
\rho_{12}(\alpha)&=\rho_{1T}\rho_{2T}+\alpha\label{E54}
\end{align}
where $0 \leq \alpha \leq \alpha_{0}$. We now define a function of the parameter $\alpha$ of a valid covariance matrix $Q(\alpha)$ as,
\begin{align}
g(\alpha)&=f_{1}(Q(\alpha))+f_{2}(Q(\alpha))+\frac{f_{1}(Q(\alpha))f_{2}(Q(\alpha))}{\left(\sigma_{Z}^{2}+\frac{\sigma_{Z_{1}}^{2}\sigma_{Z_{2}}^{2}}{(\sigma_{Z_{1}}^{2}+\sigma_{Z_{2}}^{2})}\right)}-f_{3}(Q(\alpha))\label{E55}
\end{align}
Now note the fact that
\begin{align}
 g(0)&=\frac{(1-\rho_{1T}^{2})(1-\rho_{2T}^{2})P_{1}P_{2}}{\left(\sigma_{Z}^{2}+\frac{\sigma_{Z_{1}}^{2}\sigma_{Z_{2}}^{2}}{(\sigma_{Z_{1}}^{2}+\sigma_{Z_{2}}^{2})}\right)}>0\label{E66}
 \end{align}
We are also given that $Q$ satisfies (\ref{E23}) for some
$\alpha_{0}$, which implies
 that,
 \begin{align}
 g(\alpha_{0})&<0\label{E67}
\end{align}
Now, we take the first derivative of the function $g(\alpha)$, to
obtain,
\begin{align}
\frac{d g(\alpha)}{d
\alpha}&=-2\alpha\left(\frac{P_{1}}{(1-\rho_{2T}^{2})}+\frac{P_{2}}{(1-\rho_{1T}^{2})}\right)-\frac{4\alpha
P_{1}P_{2}}{\left(\sigma_{Z}^{2}+\frac{\sigma_{Z_{1}}^{2}\sigma_{Z_{2}}^{2}}{(\sigma_{Z_{1}}^{2}+\sigma_{Z_{2}}^{2})}\right)}\left(1-\left(\frac{\alpha}{\lambda}\right)^{2}\right)-2\sqrt{P_{1}P_{2}}\nonumber\\
&\leq 0 \nonumber
\end{align}
which implies that $g(\alpha)$ is monotonically decreasing in
$\alpha$. This implies that there exists an $\alpha^{*}\in
(0,\alpha_{0})$ such that $g(\alpha^{*})=0$\footnote{We should remark here that
the existence of an $\alpha^{*}\in (0,\alpha_{0})$, with $g(\alpha^{*})=0$ can also be proved alternatively
by invoking the mean value theorem, since we have $g(0)>0$, $g(\alpha_{0})<0$ and $g(\alpha)$
is a continuous function of $\alpha$. Monotonicity of $g(\alpha)$ in fact proves a stronger statement that
such an $\alpha^{*}$ exists and is also unique.}. We use this
$\alpha^{*}$ to construct our new covariance matrix $S$ as
follows,
\begin{align}
\rho_{1T}^{(S)}&=\rho_{1T},\quad \rho_{2T}^{(S)}=\rho_{2T}\\
\rho_{12}^{(S)}&=\rho_{1T}\rho_{2T}+\alpha^{*}
\end{align}
It now remains to check wether $S$ satisfies the four conditions in
(\ref{E39})-(\ref{E42}). The condition (\ref{E42}) is met with
equality, since we have $g(\alpha^{*})=0$. Moreover,
$f_{1}(Q)=f_{1}(Q(\alpha_{0}))\leq f_{1}(Q(\alpha^{*}))=f_{1}(S)$
since $f_{1}(Q(\alpha))$ is monotonically decreasing in $\alpha$
for $\alpha \in [0,\lambda]$. Similarly, we also have
$f_{2}(Q)\leq f_{2}(S)$. Finally,
\begin{align}
f(Q)&=f_{1}(Q)+f_{2}(Q)+\frac{f_{1}(Q)f_{2}(Q)}{\left(\sigma_{Z}^{2}+\frac{\sigma_{Z_{1}}^{2}\sigma_{Z_{2}}^{2}}{(\sigma_{Z_{1}}^{2}+\sigma_{Z_{2}}^{2})}\right)}\label{E56}\\
&\leq
f_{1}(S)+f_{2}(S)+\frac{f_{1}(S)f_{2}(S)}{\left(\sigma_{Z}^{2}+\frac{\sigma_{Z_{1}}^{2}\sigma_{Z_{2}}^{2}}{(\sigma_{Z_{1}}^{2}+\sigma_{Z_{2}}^{2})}\right)}\label{E57}\\
&=f_{3}(S)
\end{align}
This shows the existence of a valid covariance matrix $S$ which
satisfies (\ref{DBNF5}) and yields a set of rates which includes the
set of rates of the given non-Gaussian distribution with the covariance matrix $Q$.

Above two cases show that for any non-Gaussian distribution
$p(t,x_{1},x_{2})$ in the set $\mathcal{P}_{NG}^{DB(b)}$, we can
always find a jointly Gaussian triple $(T_{G},X_{1G},X_{2G})$ in $\mathcal{P}_{G}^{DB}$
that yields a set of rates subsuming the set of rates of the given non-Gaussian
distribution. This consequently completes the proof of the
statement that it is sufficient to consider jointly Gaussian
$(T,X_{1},X_{2})$ in $\mathcal{P}_{G}^{DB}$ when evaluating our outer bound.

The dependence balance based outer bound can now be written in an
explicit form as follows,
\begin{align}
\mathcal{DB}_{NF}^{MAC}=\bigcup_{Q\in
\mathcal{Q}^{DB}}\Bigg\{(R_{1},R_{2}): \hspace{0.05in}&R_{1}\leq
\frac{1}{2}\mbox{log}\left(1+\frac{f_{1}(Q)}{\sigma_{Z}^{2}}\right)\nonumber\\
&R_{2}\leq \frac{1}{2}\mbox{log}\left(1+\frac{f_{2}(Q)}{\sigma_{Z}^{2}}\right)\nonumber\\
&R_{1}+R_{2}\leq \frac{1}{2}\mbox{log}\left(1+\frac{f_{3}(Q)}{\sigma_{Z}^{2}}\right)\Bigg\}\nonumber\\
\end{align}
where $\mathcal{Q}^{DB}$ is the set of $3\times 3$ covariance
matrices of the form (\ref{Qdef1}) satisfying,
\begin{align}
f_{3}(Q)&\leq f_{1}(Q)+f_{2}(Q) +
\frac{f_{1}(Q)f_{2}(Q)}{\left(\sigma_{Z}^{2}+\frac{\sigma_{Z_{1}}^{2}\sigma_{Z_{2}}^{2}}{(\sigma_{Z_{1}}^{2}+\sigma_{Z_{2}}^{2})}\right)}\label{constraint}
\end{align}
where
\begin{align}
f_{1}(Q)&=\frac{\Delta P_{1}}{(1-\rho_{2T}^{2})}\\
f_{2}(Q)&=\frac{\Delta
P_{2}}{(1-\rho_{1T}^{2})}\\
f_{3}(Q)&=(1-\rho_{1T}^{2})P_{1}+(1-\rho_{2T}^{2})P_{2}+2(\rho_{12}-\rho_{1T}\rho_{2T})\sqrt{P_{1}P_{2}}
\end{align}
and
\begin{align}
\Delta&=1-\rho_{1T}^{2}-\rho_{2T}^{2}-\rho_{12}^{2}+2\rho_{1T}\rho_{2T}\rho_{12}
\end{align}
where $\rho_{12},\rho_{1T}$ and $\rho_{2T}$ are all in $[-1,1]$.

The cut-set outer bound given in (\ref{CSMAC1})-(\ref{CSMAC4}) is evaluated for the Gaussian MAC with noisy feedback
described in (\ref{NF1})-(\ref{NF3}) as
\begin{align}
\mathcal{CS}_{NF}^{MAC}=\bigcup_{\rho\in
[0,1]}\Bigg\{(R_{1},R_{2}): \hspace{0.05in}&R_{1}\leq
\frac{1}{2}\mbox{log}\left(1+\frac{(1-\rho^{2})P_{1}}{\sigma_{Z}^{2}}\right)\nonumber\\
&R_{2}\leq \frac{1}{2}\mbox{log}\left(1+\frac{(1-\rho^{2})P_{2}}{\sigma_{Z}^{2}}\right)\nonumber\\
&R_{1}+R_{2}\leq \frac{1}{2}\mbox{log}\left(1+\frac{P_{1}+P_{2}+2\rho\sqrt{P_{1}P_{2}}}{\sigma_{Z}^{2}}\right)\Bigg\}
\end{align}

We briefly mention what our outer bound gives for the the two
limiting values of the backward noise variances
$\sigma_{Z_{1}}^{2}$ and $\sigma_{Z_{2}}^{2}$.
\begin{enumerate}
  \item $\sigma_{Z_{1}}^{2},\sigma_{Z_{2}}^{2}\rightarrow 0$ : this case corresponds to the Gaussian MAC with noiseless feedback and the constraint (\ref{constraint})
  simplifies to
  \begin{align}
  f_{3}(Q)&\leq f_{1}(Q)+f_{2}(Q)+\frac{f_{1}(Q)f_{2}(Q)}{\sigma_{Z}^{2}}
  \end{align}
  which is simply stating that the sum-rate constraint should be at most as large as the sum of the individual rate constraints, i.e., another equivalent way of writing is
  \begin{align}
  \frac{1}{2}\mbox{log}\left(1+\frac{f_{3}(Q)}{\sigma_{Z}^{2}}\right)&\leq \frac{1}{2}\mbox{log}\left(1+\frac{f_{1}(Q)}{\sigma_{Z}^{2}}\right)+\frac{1}{2}\mbox{log}\left(1+\frac{f_{2}(Q)}{\sigma_{Z}^{2}}\right)
  \end{align}
  This is the same constraint as obtained by Ozarow in \cite{OZ:1984}, and our outer bound coincides with the cut-set bound and yields the capacity region of the Gaussian MAC with noiseless feedback.
  \item $\sigma_{Z_{1}}^{2},\sigma_{Z_{2}}^{2}\rightarrow \infty$: this case corresponds to very noisy feedback and our outer bound should collapse to the no-feedback capacity region
  of the Gaussian MAC. For this case, the constraint (\ref{constraint}) simplifies to,
  \begin{align}
  f_{3}(Q)&\leq f_{1}(Q)+f_{2}(Q)
  \end{align}
  On substituting the values of $f_{1}(Q),f_{2}(Q)$ and $f_{3}(Q)$ in the above inequality, we obtain
  \begin{align}
  (\rho_{12}-\rho_{1T}\rho_{2T})&\leq \frac{((1-\rho_{1T}^{2})P_{1}+(1-\rho_{2T}^{2})P_{2})}{2\sqrt{P_{1}P_{2}}}\left(\frac{\Delta}{\lambda^{2}}-1\right)\\
  &\leq 0
  \end{align}
  where the last inequality comes from the fact that for any valid covariance matrix, $\Delta\leq \lambda^{2}$. This implies that the dependence balance bound only
  allows such covariance matrices $Q$ for which $\rho_{12}\leq \rho_{1T}\rho_{2T}$. But we know already from (\ref{E45})-(\ref{E46}) that we can always find
  an $S$ for which we can select
  $\rho_{12}^{(S)}=\rho_{1T}\rho_{2T}$, which satisfies the dependence balance bound and yields larger rates than any $Q$ with $\rho_{12}<\rho_{1T}\rho_{2T}$. Thus,
  we only need to restrict our attention to those matrices $Q$ for which $\rho_{12}=\rho_{1T}\rho_{2T}$. Such covariance matrices $Q$ correspond to those jointly
  Gaussian triples which satisfy the Markov chain $X_{1}\rightarrow T \rightarrow X_{2}$. This can be observed by noting that for any jointly Gaussian
  $(T,X_{1},X_{2})$, with a covariance matrix $Q$, the condition $I(X_{1};X_{2}|T)=0$ holds iff $\mbox{Var}(X_{1}|T)=\mbox{Var}(X_{1}|X_{2},T)$, which is equivalent to $\rho_{12}=\rho_{1T}\rho_{2T}$. Proof of this statement is immediate by noting that for a jointly Gaussian
  triple, we have
  \begin{align}
  I(X_{1};X_{2}|T)=\frac{1}{2}\mbox{log}\left(\frac{\mbox{Var}(X_{1}|T)}{\mbox{Var}(X_{1}|X_{2},T)}\right)\label{argumentNF}
 \end{align}
  Therefore, $T$ can be interpreted simply as a
  timesharing random variable and our outer bound yields the capacity region of the Gaussian MAC without feedback.
\end{enumerate}

Figure $7$ illustrates $\mathcal{DB}_{NF}^{MAC}$, the cut-set
bound and the capacity region without feedback for the cases when
$\sigma_{Z_{1}}^{2}=\sigma_{Z_{2}}^{2}=2$, $5$ and $10$, where
$P_{1}=P_{2}=\sigma_{Z}^{2}=1$. Figure $8$ illustrates
$\mathcal{DB}_{NF}^{MAC}$, the cut-set bound, the capacity region
without feedback and an achievable rate region based on
superposition coding \cite{WVMS:1983} for the case when
$\sigma_{Z_{1}}^{2}=\sigma_{Z_{2}}^{2}=0.3$ and
$P_{1}=P_{2}=\sigma_{Z}^{2}=1$.

\subsection{Remark}
For the special case of Gaussian MAC with common, noisy feedback, where
\begin{align}
Y_{F_{1}}&=Y_{F_{2}}=Y+V
\end{align}
the evaluation of $\mathcal{DB}_{NF}^{MAC}$ follows in a similar manner as in the case
of different noisy feedback signals. The only difference arises in the application of the EPI.
In particular, the regular EPI \cite{Cover:book}
suffices to provide a non-trivial upper bound on $I(X_{1},X_{2};Y|T)$ than the one provided by the maximum
entropy theorem \cite{Cover:book}. The remainder of the proof of evaluation of our outer bound for this channel model
follows along the same lines as the proof for different noisy feedback signals.
The final expressions of outer bounds for these two channel models only differ over the constraint (\ref{constraint}). For the case of
common, noisy feedback, the set $\mathcal{Q}^{DB}$ comprises of $3\times 3$ covariance
matrices of the form (\ref{Qdef1}) satisfying,
\begin{align}
f_{3}(Q)&\leq f_{1}(Q)+f_{2}(Q) +
\frac{f_{1}(Q)f_{2}(Q)}{(\sigma_{Z}^{2}+\sigma_{V}^{2})}\label{constraintcommon}
\end{align}
Now consider the Gaussian MAC with different noisy feedback
signals $Y_{F_{1}}$ and $Y_{F_{2}}$ at the transmitters $1$ and
$2$, respectively. If the variances of feedback noises $Z_{1}$ and
$Z_{2}$ are such that,
$\sigma_{Z_{1}}^{2}=\sigma_{Z_{2}}^{2}=\sigma_{V}^{2}$, then the
dependence balance constraint (\ref{constraint}) simplifies as
\begin{align}
f_{3}(Q)&\leq f_{1}(Q)+f_{2}(Q) +
\frac{f_{1}(Q)f_{2}(Q)}{\left(\sigma_{Z}^{2}+\frac{\sigma_{V}^{2}}{2}\right)}\label{constraintcommon1}
\end{align}
This implies that if a covariance matrix $Q$ satisfies the
constraint (\ref{constraintcommon}), then it also satisfies
(\ref{constraintcommon1}) but the converse statement may not
always be true. This means that the resulting outer bound for the
Gaussian MAC with common noisy feedback, with feedback noise
variance $\sigma_{V}^{2}$ can be strictly smaller than the
resulting outer bound for Gaussian MAC with different noisy
feedback signals, when the feedback noise variances are
$\sigma_{Z_{1}}^{2}=\sigma_{Z_{2}}^{2}=\sigma_{V}^{2}$.

\begin{figure}[p]
  \centerline{\epsfig{figure=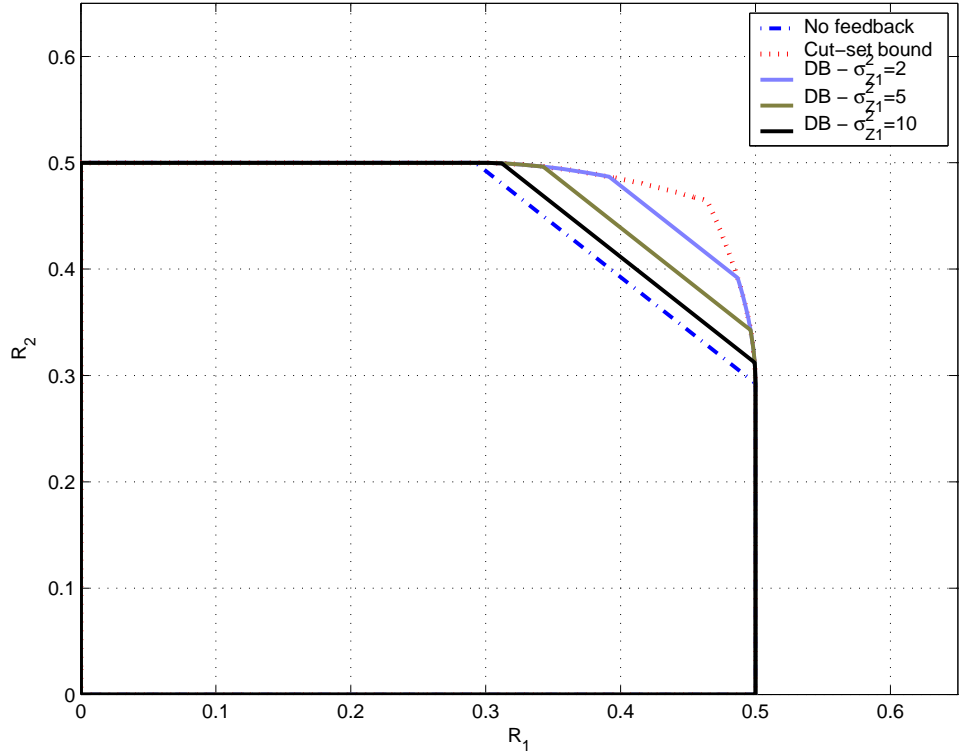,width=12.8cm}}
  \caption{Illustration of outer bounds for
  $P_{1}=P_{2}=\sigma_{Z}^{2}=1$ and
  $\sigma_{Z_{1}}^{2}=\sigma_{Z_{2}}^{2}=2,5,10$.}\label{fig7}
  \end{figure}
\begin{figure}
  \centerline{\epsfig{figure=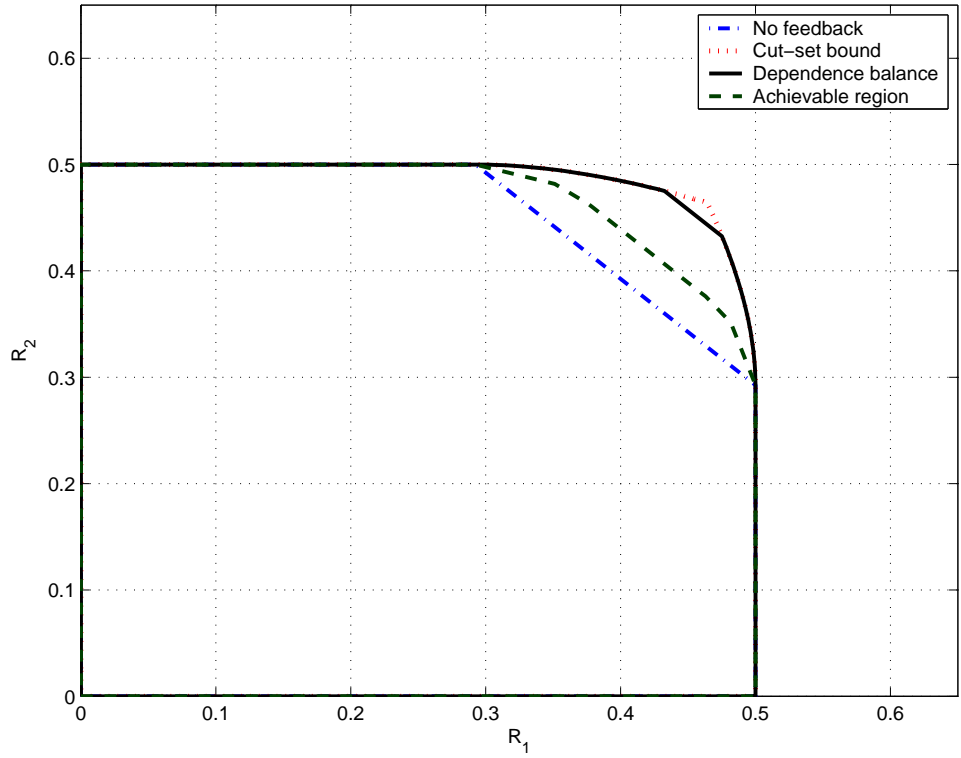,width=12.8cm}}
  \caption{Illustration of outer bound and an achievable region based on superposition coding
  for
  $P_{1}=P_{2}=\sigma_{Z}^{2}=1$ and $\sigma_{Z_{1}}^{2}=\sigma_{Z_{2}}^{2}=0.3$.}\label{fig8}
\end{figure}

\newpage
\section{Evaluation of $\mathcal{DB}_{UC}^{MAC}$}

In this section we will explicitly evaluate Theorem $4$ for the
Gaussian MAC with user cooperation described by
(\ref{UCmodel1})-(\ref{UCmodel3}) in Section $7$. We start with
step $1$ and characterize the set of jointly Gaussian triples
$(T_{G},X_{1G},X_{2G})$ in $\mathcal{P}_{G}^{DB}$. For this
purpose, we rewrite (\ref{DBT6}) as follows,
\begin{align}
0&\leq I(X_{1};X_{2}|Y_{F_{1}},Y_{F_{2}},T)-I(X_{1};X_{2}|T)\label{Rew1}\\
&= I(X_{1};Y_{F_{1}},Y_{F_{2}}|X_{2},T)-I(X_{1};Y_{F_{1}},Y_{F_{2}}|T)\label{Rew2}\\
&=
 h(Y_{F_{1}},Y_{F_{2}}|X_{1},T)+h(Y_{F_{1}},Y_{F_{2}}|X_{2},T)-h(Y_{F_{1}},Y_{F_{2}}|T)-h(Y_{F_{1}},Y_{F_{2}}|X_{1},X_{2},T)\label{Rew3}
\end{align}
and express the above constraint as follows,
\begin{align}
h(Y_{F_{1}},Y_{F_{2}}|T)+h(Y_{F_{1}},Y_{F_{2}}|X_{1},X_{2},T)&\leq
h(Y_{F_{1}},Y_{F_{2}}|X_{1},T)+h(Y_{F_{1}},Y_{F_{2}}|X_{2},T)\label{Rew4}
\end{align}
Making use of the following equalities,
\begin{align}
h(Y_{F_{1}},Y_{F_{2}}|X_{1},X_{2},T)&=\frac{1}{2}\mbox{log}\left((2\pi \mbox{e})^{2}\sigma_{Z_{1}}^{2}\sigma_{Z_{2}}^{2}\right)\\
h(Y_{F_{1}},Y_{F_{2}}|X_{1},T)&=\frac{1}{2}\mbox{log}\left((2\pi \mbox{e})\sigma_{Z_{2}}^{2}\right) + h(Y_{F_{1}}|X_{1},T)\\
h(Y_{F_{1}},Y_{F_{2}}|X_{2},T)&=\frac{1}{2}\mbox{log}\left((2\pi \mbox{e})\sigma_{Z_{1}}^{2}\right) + h(Y_{F_{2}}|X_{2},T)
\end{align}
we obtain a simplified expression for (\ref{Rew4}) as,
\begin{align}
&h(Y_{F_{1}},Y_{F_{2}}|T)\leq
h(Y_{F_{1}}|X_{1},T)+h(Y_{F_{2}}|X_{2},T)\label{Rew5}
\end{align}
We further simplify (\ref{Rew5}) as follows,
\begin{align}
0 &\leq h(Y_{F_{1}}|X_{1},T)+h(Y_{F_{2}}|X_{2},T)-h(Y_{F_{1}},Y_{F_{2}}|T)
\label{Rew6}\\
&= h(Y_{F_{1}}|X_{1},T)+h(Y_{F_{2}}|X_{2},T)-h(Y_{F_{1}}|T)-h(Y_{F_{2}}|Y_{F_{1}},T)
\label{Rew7}\\
&= -I(Y_{F_{1}};X_{1}|T)+h(Y_{F_{2}}|X_{2},T)-h(Y_{F_{2}}|Y_{F_{1}},T)
\label{Rew8}\\
&= -I(Y_{F_{1}};X_{1}|T)+h(Y_{F_{2}}|X_{2},Y_{F_{1}},T)-h(Y_{F_{2}}|Y_{F_{1}},T)
\label{Rew9}\\
&= -I(Y_{F_{1}};X_{1}|T)-I(Y_{F_{2}};X_{2}|Y_{F_{1}},T)
\label{Rew10}
\end{align}
where (\ref{Rew9}) follows from the Markov chain $Y_{F_{1}}\rightarrow X_{2} \rightarrow (T,Y_{F_{2}})$. Therefore,
the dependence balance constraint in (\ref{DBT6}) is equivalent to following two equalities,
\begin{align}
I(Y_{F_{1}};X_{1}|T)&=0\label{sim1}\\
I(Y_{F_{2}};X_{2}|Y_{F_{1}},T)&=0\label{sim2}
\end{align}
Next, we show that if any jointly Gaussian triple $(T,X_{1},X_{2})$ satisfies the constraints (\ref{sim1})-(\ref{sim2})
then it satisfies the Markov chain $X_{1}\rightarrow T \rightarrow X_{2}$. Conversely, we will show that if
any jointly Gaussian triple $(T,X_{1},X_{2})$ satisfies $X_{1}\rightarrow T \rightarrow X_{2}$, then it satisfies (\ref{sim1})-(\ref{sim2}).

We start by evaluating (\ref{sim1}) and (\ref{sim2}) for a jointly Gaussian $(T_{G},X_{1G},X_{2G})$ which is equivalent to,
\begin{align}
0&=I(\sqrt{h_{21}}X_{2G}+Z_{1};X_{1G}|T_{G})\label{Lem2}\\
0&=I(\sqrt{h_{12}}X_{1G}+Z_{2};X_{2G}|\sqrt{h_{21}}X_{2G}+Z_{1},T_{G})\label{Lem3}
\end{align}
These equalities are equivalent to
\begin{align}
\mbox{Cov}(X_{1G},X_{2G}|T_{G})&=0\label{Lem5}
\end{align}
Using the same argument as in (\ref{argumentNF}), we obtain the following condition
\begin{align}
\rho_{12}=\rho_{1T}\rho_{2T}\label{Lem9}
\end{align}
This implies that a jointly Gaussian triple satisfies (\ref{sim1})-(\ref{sim2}) iff $\rho_{12}=\rho_{1T}\rho_{2T}$.

On the other hand, consider any jointly Gaussian triple $(T_{G},X_{1G},X_{2G})$, with a covariance matrix $Q$
which satisfies the Markov chain $X_{1G}\rightarrow T_{G} \rightarrow X_{2G}$.
This is equivalent to $I(X_{1G};X_{2G}|T_{G})=0$,
which is equivalent to
\begin{align}
\rho_{12}&=\rho_{1T}\rho_{2T}\label{lemtem}
\end{align}
This implies that if a jointly Gaussian triple $(T,X_{1},X_{2})$ satisfies the Markov chain $X_{1}\rightarrow T \rightarrow X_{2}$, then it satisfies
(\ref{lemtem}) and therefore it also satisfies (\ref{sim1})-(\ref{sim2}) and vice versa.
As a consequence, we have explicitly characterized the set $\mathcal{P}_{G}^{DB}$, i.e., it comprises of
only such jointly Gaussian distributions, $(T_{G},X_{1G},X_{2G})$, for which $X_{1G}\rightarrow T_{G} \rightarrow X_{2G}$.

We can now write the set of rate pairs provided by our outer bound for a jointly Gaussian triple $(T_{G},X_{1G},X_{2G})$ in the set $\mathcal{P}_{G}^{DB}$ as
\begin{align}
R_{1} &\leq I(X_{1G};Y,Y_{F_{2}}|X_{2G},T_{G})\\
R_{2} &\leq I(X_{2G};Y,Y_{F_{1}}|X_{1G},T_{G})\\
R_{1}+R_{2} &\leq I(X_{1G},X_{2G};Y,Y_{F_{1}},Y_{F_{2}}|T_{G})\\
R_{1}+R_{2}&\leq I(X_{1G},X_{2G};Y)
\end{align}
where $(T_{G},X_{1G},X_{2G})$ satisfies the Markov chain
$X_{1G}\rightarrow T_{G} \rightarrow X_{2G}$. Moreover, from the
evaluation of step $2$ in Section $9$, we know that all rate pairs
contributed by input distributions in $\mathcal{P}_{NG}^{{DB(a)}}$
are covered by those given in $\mathcal{P}_{G}^{DB}$. Therefore,
we do not need to consider the set $\mathcal{P}_{NG}^{{DB(a)}}$ in
evaluating our outer bound.

We now arrive at step $3$ of the evaluation of our outer bound
where we will show that for any non-Gaussian input distribution
$p(t,x_{1},x_{2}) \in \mathcal{P}_{NG}^{DB(b)}$, we can always
find an input distribution in $\mathcal{P}_{G}^{DB}$, with a set
of rate pairs which include the set of rate pairs of the fixed
non-Gaussian input distribution $p(t,x_{1},x_{2})$. Consider any
triple $(T,X_{1},X_{2})$ with a non-Gaussian input distribution
$p(t,x_{1},x_{2})\in \mathcal{P}_{NG}^{DB(b)}$, with a valid
covariance matrix $Q$. By the definition of the set
$\mathcal{P}_{NG}^{DB(b)}$, and as a consequence of (\ref{Lem9}),
this covariance matrix has the property that $\rho_{12}\neq
\rho_{1T}\rho_{2T}$. Moreover, this non-Gaussian distribution
satisfies the dependence balance bound, i.e., it satisfies
(\ref{sim1}) and (\ref{sim2}). For our purpose, we only need
(\ref{sim1}). Since $I(Y_{F_{1}};X_{1}|T)=0$, this implies
\begin{align}
E[(\sqrt{h_{21}}X_{2}+Z_{1})X_{1}|T]&=E[\sqrt{h_{21}}X_{2}+Z_{1}|T]E[X_{1}|T]\label{Exp1}\\
&=\sqrt{h_{21}}E[X_{2}|T]E[X_{1}|T]\label{Exp2}
\end{align}
on the other hand, we also have $E[(\sqrt{h_{21}}X_{2}+Z_{1})X_{1}|T]=\sqrt{h_{21}}E[X_{1}X_{2}|T]$, which implies
\begin{align}
E[X_{1}X_{2}|T]&=E[X_{2}|T]E[X_{1}|T]\label{Exp3}
\end{align}

We will now construct another triple $(T^{'},X_{1},X_{2})$ with a covariance
matrix $S$ by selecting
\begin{align}
T^{'}&=E[X_{1}|T]\label{TP1}
\end{align}
This particular selection is closely related to the recent work of Bross, Lapidoth and Wigger \cite{LapidothISIT:2008}
where it was shown that jointly Gaussian distributions are sufficient to characterize the capacity region
of Gaussian MAC with conferencing encoders. Although, we should also remark that when evaluating our outer bound for user cooperation, we do not have a conditionally
independent structure among $(T,X_{1},X_{2})$ to start with. This structure arises from the dependence balance constraint (\ref{DBT6}),
permiting us to use this approach.

Returning to (\ref{TP1}), we note that $T^{'}$ is a deterministic function of $T$ and therefore,
following is a valid Markov chain.
\begin{align}
T^{'}\rightarrow T\rightarrow (X_{1},X_{2}) \rightarrow (Y,Y_{F_{1}},Y_{F_{2}})\label{TP2}
\end{align}
We will now obtain the off diagonal elements of the covariance matrix $S$ of the triple $(T^{'},X_{1},X_{2})$ as follows,
\begin{align}
E[X_{1}T^{'}]&=E_{T}[E[X_{1}T^{'}|T]]\label{TP4}\\
&=E_{T}[E[X_{1}|T]E[X_{1}|T]]\label{TP5}\\
&=\mbox{Var}(T^{'})\label{TP6}
\end{align}
and
\begin{align}
E[X_{2}T^{'}]&=E_{T}[E[X_{2}T^{'}|T]]\label{TP8}\\
&=E_{T}[E[X_{2}|T]E[X_{1}|T]]\label{TP9}
\end{align}
and finally,
\begin{align}
E[X_{1}X_{2}]&=E_{T}[E[X_{1}X_{2}|T]]\label{TP11}\\
&=E_{T}[E[X_{1}|T]E[X_{2}|T]]\label{TP12}
\end{align}
where (\ref{TP12}) follows from (\ref{Exp3}). Therefore, the triple
$(T^{'},X_{1},X_{2})$ satisfies
\begin{align}
E[X_{1}X_{2}]&=\frac{E[X_{1}T^{'}]E[X_{2}T^{'}]}{\mbox{Var}(T^{'})}\label{TP13}
\end{align}
Now using the fact that
\begin{align}
E[X_{1}X_{2}]&=\rho_{12}\sqrt{P_{1}P_{2}}\label{TP14}\\
E[X_{1}T^{'}]&=\rho_{1T^{'}}\sqrt{P_{1}P_{T^{'}}}\label{TP15}\\
E[X_{2}T^{'}]&=\rho_{2T^{'}}\sqrt{P_{2}P_{T^{'}}}\label{TP16}
\end{align}
and substituting in (\ref{TP13}) we obtain that the covariance matrix $S$ satisfies
\begin{align}
\rho_{12}&=\rho_{1T^{'}}\rho_{2T^{'}}\label{TP17}
\end{align}
Therefore, from (\ref{Lem9}) any jointly Gaussian $(T^{'}_{G},X_{1G},X_{2G})$ triple with a
covariance matrix $S$, with entries $(\rho_{12},\rho_{1T^{'}},\rho_{2T^{'}})$ satisfies (\ref{DBT6}).

We now arrive at the final step of the evaluation.
In particular, we will show that the rates of this jointly Gaussian triple $(T^{'}_{G},X_{1G},X_{2G})$
will include the rates of the given non-Gaussian triple $(T,X_{1},X_{2})$. For the triple $(T^{'}_{G},X_{1G},X_{2G})$,
we have the following set of inequalities,
\begin{align}
I(X_{1G};Y,Y_{F_{2}}|X_{2G},T^{'}_{G})
&=h(Y,Y_{F_{2}}|X_{2G},T^{'}_{G})-h(Y,Y_{F_{2}}|X_{1G},X_{2G},T^{'}_{G})\label{ineq1}\\
&=h(\sqrt{h_{10}}X_{1G}+Z,\sqrt{h_{12}}X_{1G}+Z_{2}|X_{2G},T^{'}_{G})\nonumber\\&\hspace{0.17in}-h(Y,Y_{F_{2}}|X_{1G},X_{2G},T^{'}_{G})\label{ineq2}\\
&\geq h(\sqrt{h_{10}}X_{1}+Z,\sqrt{h_{12}}X_{1}+Z_{2}|X_{2},T^{'})-h(Y,Y_{F_{2}}|X_{1G},X_{2G},T^{'}_{G})\label{ineq3}\\
&\geq h(\sqrt{h_{10}}X_{1}+Z,\sqrt{h_{12}}X_{1}+Z_{2}|X_{2},T^{'},T)-h(Y,Y_{F_{2}}|X_{1},X_{2},T)\label{ineq4}\\
&= h(\sqrt{h_{10}}X_{1}+Z,\sqrt{h_{12}}X_{1}+Z_{2}|X_{2},T)-h(Y,Y_{F_{2}}|X_{1},X_{2},T)\label{ineq5}\\
&=I(X_{1};Y,Y_{F_{2}}|X_{2},T)\label{ineq6}
\end{align}
where (\ref{ineq3}) follows from the fact that $(T^{'},X_{1},X_{2})$ and $(T^{'}_{G},X_{1G},X_{2G})$ have the same covariance matrix $S$ and by
using the maximum entropy theorem. Next, (\ref{ineq4}) follows from the fact that conditioning reduces differential entropy and finally (\ref{ineq5})
follows from the fact that $T^{'}$ is a deterministic function of $T$ and by invoking the Markov chain in (\ref{TP2}). Similarly, we also have
\begin{align}
I(X_{2G};Y,Y_{F_{1}}|X_{1G},T^{'}_{G})&\geq I(X_{2};Y,Y_{F_{1}}|X_{1},T)\label{ineq7}\\
I(X_{1G},X_{2G};Y,Y_{F_{1}},Y_{F_{2}}|T^{'}_{G})&\geq I(X_{1},X_{2};Y,Y_{F_{1}},Y_{F_{2}}|T)\label{ineq8}
\end{align}
Finally, we have
\begin{align}
I(X_{1G},X_{2G};Y)&=h(Y)-h(Y|X_{1G},X_{2G})\label{ineq9}\\
&=h(\sqrt{h_{10}}X_{1G}+\sqrt{h_{20}}X_{2G}+Z)-h(Z)\label{ineq10}\\
&\geq h(\sqrt{h_{10}}X_{1}+\sqrt{h_{20}}X_{2}+Z)-h(Z)\label{ineq11}\\
&=I(X_{1},X_{2};Y)\label{ineq13}
\end{align}
Therefore, we conclude that for any non-Gaussian distribution
$p(t,x_{1},x_{2})\in \mathcal{P}_{NG}^{DB(b)}$, there exists a
jointly Gaussian distribution $p(t,x_{1},x_{2})\in
\mathcal{P}_{G}^{DB}$ which satisfies the dependence balance bound
(\ref{DBT6}) and yields a set of rates which include the set of
rates given by the fixed non-Gaussian distribution. Hence, it
suffices to consider jointly Gaussian distributions in
$\mathcal{P}_{G}^{DB}$ to evaluate our outer bound.

The dependence balance based outer bound can now be written in an
explicit form as follows,
\begin{align}
\mathcal{DB}_{UC}^{MAC}=\bigcup_{(\rho_{1T},\rho_{2T}) \in [0,1]\times
[0,1]}\Bigg\{(R_{1},R_{2}): \hspace{0.05in}&R_{1}\leq
\frac{1}{2}\mbox{log}\left(1+f_{1}(\rho_{1T})\right)\nonumber\\
&R_{2}\leq \frac{1}{2}\mbox{log}\left(1+f_{2}(\rho_{2T})\right)\nonumber\\
&R_{1}+R_{2}\leq \frac{1}{2}\mbox{log}\left(1+f_{3}(\rho_{1T},\rho_{2T})\right)\nonumber\\
&R_{1}+R_{2}\leq \frac{1}{2}\mbox{log}\left(1+f_{4}(\rho_{1T},\rho_{2T})\right)
\Bigg\}
\end{align}
where
\begin{align}
f_{1}(\rho_{1T})&=(1-\rho_{1T}^{2})P_{1}\left(\frac{h_{10}}{\sigma_{Z}^{2}}+\frac{h_{12}}{\sigma_{Z_{2}}^{2}}\right)\\
f_{2}(\rho_{2T})&=(1-\rho_{2T}^{2})P_{2}\left(\frac{h_{20}}{\sigma_{Z}^{2}}+\frac{h_{21}}{\sigma_{Z_{1}}^{2}}\right)\\
f_{3}(\rho_{1T},\rho_{2T})&=f_{1}(\rho_{1T})+f_{2}(\rho_{2T})+(1-\rho_{1T}^{2})(1-\rho_{2T}^{2})P_{1}P_{2}\beta\\
f_{4}(\rho_{1T},\rho_{2T})&=\frac{(h_{10}P_{1}+h_{20}P_{2}+2\rho_{1T}\rho_{2T}\sqrt{h_{10}h_{20}P_{1}P_{2}})}{\sigma_{Z}^{2}}
\end{align}
and
\begin{align}
\beta &=\frac{(h_{12}h_{21}\sigma_{Z}^{2}+h_{20}h_{12}\sigma_{Z_{1}}^{2}+h_{10}h_{21}\sigma_{Z_{2}}^{2})}{\sigma_{Z}^{2}\sigma_{Z_{1}}^{2}\sigma_{Z_{2}}^{2}}
\end{align}

The cut-set outer bound given in (\ref{CSMAC1})-(\ref{CSMAC4}) is evaluated for the Gaussian MAC with user
cooperation described in (\ref{UCmodel1})-(\ref{UCmodel3}) as
\begin{align}
\mathcal{CS}_{UC}^{MAC}=\bigcup_{\rho \in [0,1]}\Bigg\{(R_{1},R_{2}):
\hspace{0.05in}&R_{1}\leq
\frac{1}{2}\mbox{log}\left(1+(1-\rho^{2})P_{1}\left(\frac{h_{10}}{\sigma_{Z}^{2}}+\frac{h_{12}}{\sigma_{Z_{2}}^{2}}\right)\right)\nonumber\\
&R_{2}\leq \frac{1}{2}\mbox{log}\left(1+(1-\rho^{2})P_{2}\left(\frac{h_{20}}{\sigma_{Z}^{2}}+\frac{h_{21}}{\sigma_{Z_{1}}^{2}}\right)\right)\nonumber\\
&R_{1}+R_{2}\leq \frac{1}{2}\mbox{log}\left(1+\frac{h_{10}P_{1}+h_{20}P_{2}+2\rho\sqrt{h_{10}h_{20}P_{1}P_{2}}}{\sigma_{Z}^{2}}\right)
\Bigg\}
\end{align}

We now mention how our outer bound compares with the cut-set bound for the limiting cases of cooperation noise variances.
\begin{enumerate}
\item $\sigma_{Z_{1}}^{2},\sigma_{Z_{2}}^{2}\rightarrow 0$: this case corresponds to total cooperation between transmitters.
In this case, both dependence balance bound and the cut-set bound
degenerate to the total cooperation line,
\begin{align}
R_{1}+R_{2}&\leq \frac{1}{2}\mbox{log}\left(1+\frac{h_{10}P_{1}+h_{20}P_{2}+2\sqrt{h_{10}h_{20}P_{1}P_{2}}}{\sigma_{Z}^{2}}\right)
\end{align}

\item $\sigma_{Z_{1}}^{2},\sigma_{Z_{2}}^{2}\rightarrow \infty$: this case corresponds to very noisy cooperation links. In this case, we have
\begin{align}
f_{1}(\rho_{1T})&=\frac{(1-\rho_{1T}^{2})h_{10}P_{1}}{\sigma_{Z}^{2}}\\
f_{2}(\rho_{2T})&=\frac{(1-\rho_{2T}^{2})h_{20}P_{2}}{\sigma_{Z}^{2}}\\
f_{3}(\rho_{1T},\rho_{2T})&=f_{1}(\rho_{1T})+f_{2}(\rho_{2T})\\
&< \frac{(h_{10}P_{1}+h_{20}P_{2}+2\rho_{1T}\rho_{2T}\sqrt{h_{10}h_{20}P_{1}P_{2}})}{\sigma_{Z}^{2}}
\end{align}
and the dependence balance bound collapses to the capacity region of the Gaussian MAC with no cooperation.
On the other hand, the cut-set bound collapses to the capacity region of the Gaussian MAC with noiseless feedback \cite{OZ:1984}.
\end{enumerate}

Figure $9$ illustrates the outer bounds and achievable rate region
\cite{Aazhang:2003} for the case when $P_{1}=P_{2}=5,
\sigma_{Z}^{2}=2$ and $\sigma_{Z_{1}}^{2}=\sigma_{Z_{2}}^{2}=1$
and $h_{10}=h_{20}=h_{12}=h_{21}=1$. Figure $10$ illustrates the
outer bounds for the case when $P_{1}=P_{2}=\sigma_{Z}^{2}=1$ and
$\sigma_{Z_{1}}^{2}=\sigma_{Z_{2}}^{2}=20$ and
$h_{10}=h_{20}=h_{12}=h_{21}=1$. For this case, the achievable
rate region does not provide any visual improvement over
no-cooperation. Figure $11$ illustrates these bounds and the
achievable rate region for the asymmetric setting where
$P_{1}=P_{2}=\sigma_{Z}^{2}=1$ and
$\sigma_{Z_{1}}^{2}=\sigma_{Z_{2}}^{2}=1$ and $h_{10}=h_{20}=1$,
$h_{12}=3, h_{21}=2$. Figure $12$ illustrates these bounds and the
achievable rate region for the one sided cooperation where
$P_{1}=P_{2}=\sigma_{Z}^{2}=1$ and
$\sigma_{Z_{1}}^{2}=\sigma_{Z_{2}}^{2}=1$ and $h_{10}=h_{20}=1$,
$h_{12}=2, h_{21}=0$.

\begin{figure}[p]
  \centerline{\epsfig{figure=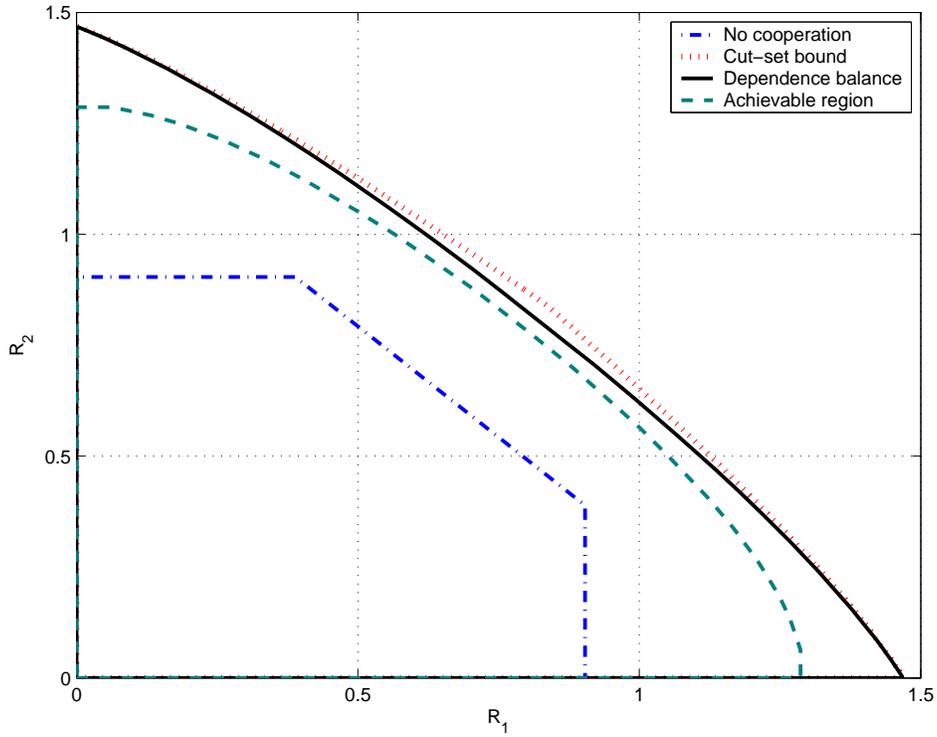,width=12.5cm}}
  \caption{Illustration of bounds for $P_{1}=P_{2}=5, \sigma_{Z}^{2}=2$, $\sigma_{Z_{1}}^{2}=\sigma_{Z_{2}}^{2}=1$ and $h_{10}=h_{20}=h_{12}=h_{21}=1$.}\label{fig9}
\end{figure}
\begin{figure}
  \centerline{\epsfig{figure=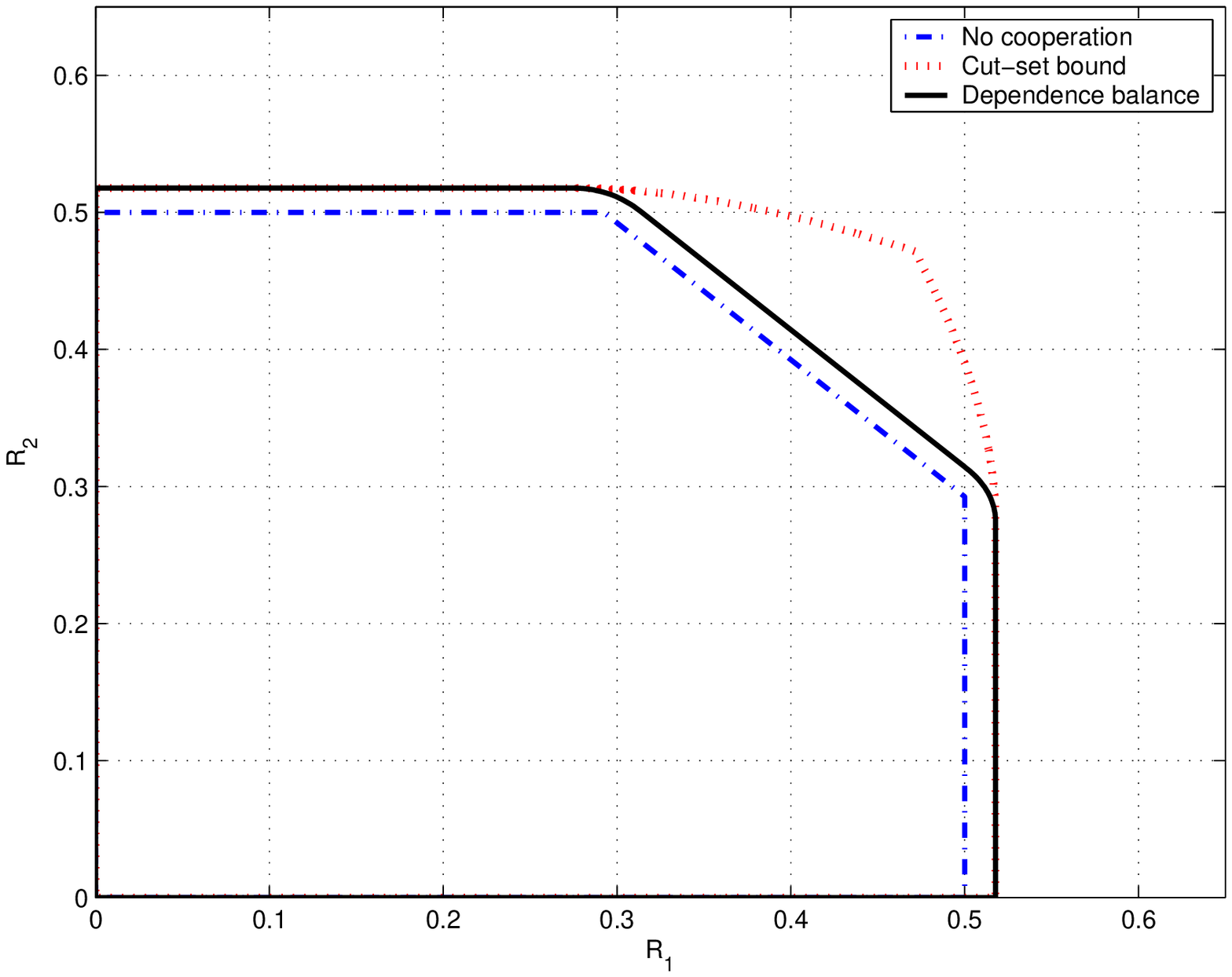,width=12.5cm}}
  \caption{Illustration of outer bounds for $P_{1}=P_{2}=\sigma_{Z}^{2}=1$, $\sigma_{Z_{1}}^{2}=\sigma_{Z_{2}}^{2}=20$ and $h_{10}=h_{20}=h_{12}=h_{21}=1$.}\label{fig10}
\end{figure}

\begin{figure}[p]
\centerline{\epsfig{figure=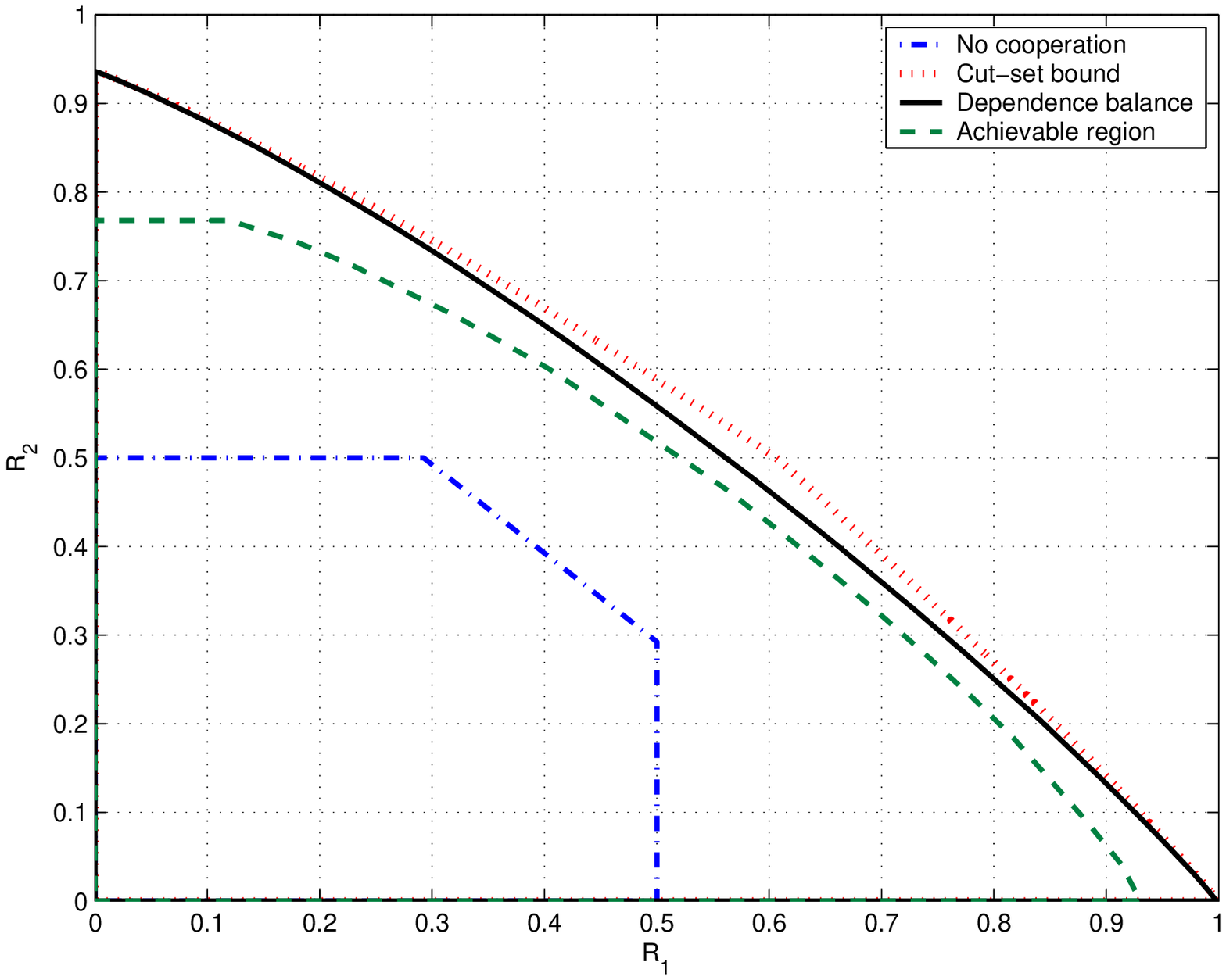,width=12.5cm}}
  \caption{Illustration of outer bounds for $P_{1}=P_{2}=\sigma_{Z}^{2}=1$, $\sigma_{Z_{1}}^{2}=\sigma_{Z_{2}}^{2}=1$ and $h_{10}=h_{20}=1$, $h_{12}=3, h_{21}=2$.}\label{fig11}
\end{figure}
\begin{figure}
\centerline{\epsfig{figure=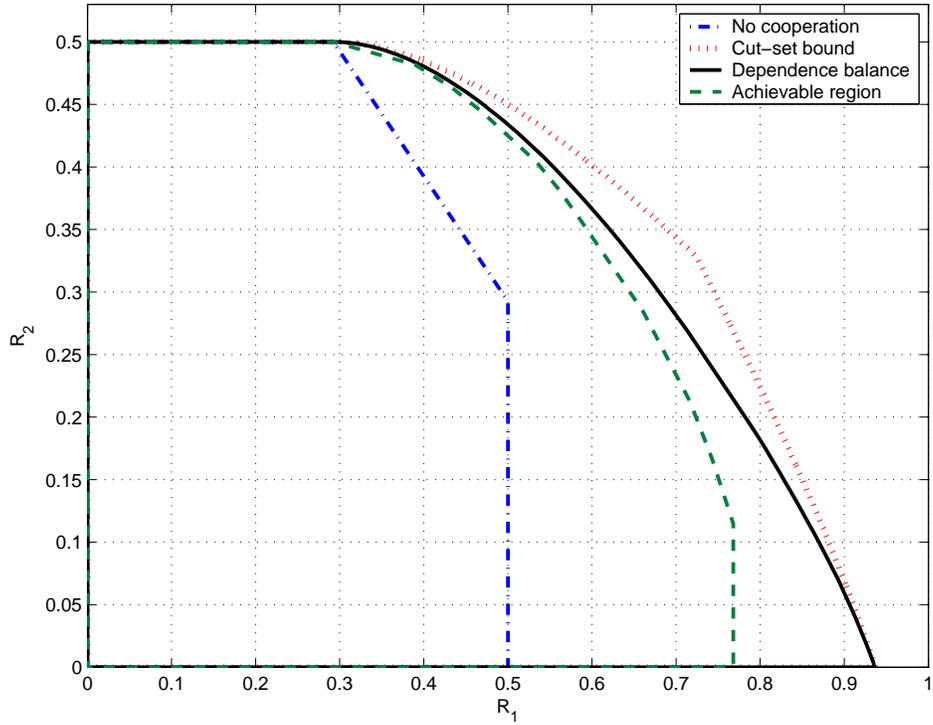,width=12.5cm}}
  \caption{Illustration of outer bounds for $P_{1}=P_{2}=\sigma_{Z}^{2}=1$, $\sigma_{Z_{1}}^{2}=\sigma_{Z_{2}}^{2}=1$ and $h_{10}=h_{20}=1$, $h_{12}=2, h_{21}=0$.}\label{fig12}
\end{figure}
\newpage
\section{Evaluation of $\mathcal{DB}_{UC}^{IC}$}

In this section we will explicitly evaluate Theorem $5$ for the
Gaussian IC with user cooperation described by
(\ref{ICUCmodel1})-(\ref{ICUCmodel4}) in Section $8$. We start
with step $1$ and first characterize the set of jointly Gaussian
triples $(T_{G},X_{1G},X_{2G})$ in $\mathcal{P}_{G}^{DB}$. For
this purpose, we rewrite (\ref{DBICT8}) as follows,
\begin{align}
h(Y_{F_{1}},Y_{F_{2}}|T)+h(Y_{F_{1}},Y_{F_{2}}|X_{1},X_{2},T)&\leq
h(Y_{F_{1}},Y_{F_{2}}|X_{1},T)+h(Y_{F_{1}},Y_{F_{2}}|X_{2},T)\label{ICRew4}
\end{align}
Making use of the following equalities,
\begin{align}
h(Y_{F_{1}},Y_{F_{2}}|X_{1},X_{2},T)&=\frac{1}{2}\mbox{log}\left((2\pi \mbox{e})^{2}\sigma_{Z_{1}}^{2}\sigma_{Z_{2}}^{2}\right)\\
h(Y_{F_{1}},Y_{F_{2}}|X_{1},T)&=\frac{1}{2}\mbox{log}\left((2\pi \mbox{e})\sigma_{Z_{2}}^{2}\right) + h(Y_{F_{1}}|X_{1},T)\\
h(Y_{F_{1}},Y_{F_{2}}|X_{2},T)&=\frac{1}{2}\mbox{log}\left((2\pi \mbox{e})\sigma_{Z_{1}}^{2}\right) + h(Y_{F_{2}}|X_{2},T)
\end{align}
we obtain a simplified expression for (\ref{ICRew4}) as,
\begin{align}
&h(Y_{F_{1}},Y_{F_{2}}|T)\leq
h(Y_{F_{1}}|X_{1},T)+h(Y_{F_{2}}|X_{2},T)\label{ICRew5}
\end{align}
which can be further simplified as in the derivation of $\mathcal{DB}_{UC}^{MAC}$
to the following two equalities,
\begin{align}
I(Y_{F_{1}};X_{1}|T)&=0\label{ICsim1}\\
I(Y_{F_{2}};X_{2}|Y_{F_{1}},T)&=0\label{ICsim2}
\end{align}
We next follow the same set of arguments used in Section $11$ to arrive at the fact that a jointly Gaussian triple $(T,X_{1},X_{2})$
satisfies (\ref{ICsim1})-(\ref{ICsim2}) iff $X_{1}\rightarrow T \rightarrow X_{2}$.

We can now write the set of rate pairs provided by our outer bound for a jointly Gaussian triple $(T_{G},X_{1G},X_{2G})$ in the set $\mathcal{P}_{G}^{DB}$ as
\begin{align}
R_{1}&\leq I(X_{1G},X_{2G};Y_{1})\\
R_{2}&\leq I(X_{1G},X_{2G};Y_{2})\\
R_{1}&\leq I(X_{1G};Y_{1},Y_{2},Y_{F_{2}}|X_{2G},T)\\
R_{2}&\leq I(X_{2G};Y_{1},Y_{2},Y_{F_{1}}|X_{1G},T)\\
R_{1}+R_{2}&\leq I(X_{1G},X_{2G};Y_{1},Y_{2},Y_{F_{1}},Y_{F_{2}}|T)\\
R_{1}+R_{2}&\leq I(X_{1G},X_{2G};Y_{1},Y_{2})\big\}
\end{align}
where the triple $(T_{G},X_{1G},X_{2G})$ satisfies the Markov
chain $X_{1G}\rightarrow T_{G} \rightarrow X_{2G}$. Moreover, from
the evaluation of step $2$ in Section $9$, we know that all rate
pairs contributed by input distributions in
$\mathcal{P}_{NG}^{{DB(a)}}$ are covered by those given in
$\mathcal{P}_{G}^{DB}$. Therefore, we do not need to consider the
set $\mathcal{P}_{NG}^{{DB(a)}}$ in evaluating our outer bound.

We now arrive at step $3$ of the evaluation of our outer bound for
the Gaussian IC with user cooperation. Consider any triple
$(T,X_{1},X_{2})$ with a non-Gaussian distribution
$p(t,x_{1},x_{2})\in \mathcal{P}_{NG}^{DB(b)}$, with a valid
covariance matrix $Q$. As in the derivation of
$\mathcal{DB}_{UC}^{MAC}$, we first construct another triple
$(T^{'},X_{1},X_{2})$ with a covariance matrix $S$ by selecting
\begin{align}
T^{'}&=E[X_{1}|T]\label{ICTP1}
\end{align}
Following this step, we next make use of the Markov chain
\begin{align}
T^{'}\rightarrow T\rightarrow (X_{1},X_{2}) \rightarrow (Y_{1},Y_{2},Y_{F_{1}},Y_{F_{2}})\label{ICTP2}
\end{align}
to show the existence of a jointly Gaussian $(T^{'}_{G},X_{1G},X_{2G})$ with a
covariance matrix $S$ and which satisfies (\ref{DBICT8}).

We now arrive at the final step of the evaluation.
In particular, we will show that the rates of this jointly Gaussian triple $(T^{'}_{G},X_{1G},X_{2G})$
will include the rates of the given non-Gaussian triple $(T,X_{1},X_{2})$. For the triple $(T^{'}_{G},X_{1G},X_{2G})$,
we have the following set of inequalities,
\begin{align}
I(X_{1G};Y_{1},Y_{2},Y_{F_{2}}|X_{2G},T^{'}_{G})
&=h(Y_{1},Y_{2},Y_{F_{2}}|X_{2G},T^{'}_{G})-h(Y_{1},Y_{2},Y_{F_{2}}|X_{1G},X_{2G},T^{'}_{G})\label{ICineq1}\\
&=h(X_{1G}+N_{1},\sqrt{a}X_{1G}+N_{2},\sqrt{h_{12}}X_{1G}+Z_{2}|X_{2G},T^{'}_{G})\nonumber\\
&\hspace{0.18in}-h(Y_{1},Y_{2},Y_{F_{2}}|X_{1G},X_{2G},T^{'}_{G})\label{ICineq2}\\
&\geq h((X_{1}+N_{1},\sqrt{a}X_{1}+N_{2},\sqrt{h_{12}}X_{1}+Z_{2}|X_{2},T^{'})\nonumber\\
&\hspace{0.18in}-h(Y_{1},Y_{2},Y_{F_{2}}|X_{1G},X_{2G},T^{'}_{G})\label{ICineq3}\\
&\geq h((X_{1}+N_{1},\sqrt{a}X_{1}+N_{2},\sqrt{h_{12}}X_{1}+Z_{2}|X_{2},T^{'},T)\nonumber\\
&\hspace{0.18in}-h(Y_{1},Y_{2},Y_{F_{2}}|X_{1G},X_{2G},T^{'}_{G})\label{ICineq4}\\
&= h((X_{1}+N_{1},\sqrt{a}X_{1}+N_{2},\sqrt{h_{12}}X_{1}+Z_{2}|X_{2},T)\nonumber\\
&\hspace{0.18in}-h(Y_{1},Y_{2},Y_{F_{2}}|X_{1},X_{2},T)\label{ICineq5}\\
&=I(X_{1};Y_{1},Y_{2},Y_{F_{2}}|X_{2},T)\label{ICineq6}
\end{align}
where (\ref{ICineq3}) follows from the fact that $(T^{'},X_{1},X_{2})$ and $(T^{'}_{G},X_{1G},X_{2G})$ have the same covariance matrix $S$ and using the
maximum entropy theorem. Next, (\ref{ICineq4}) follows from the fact that conditioning reduces differential entropy and finally (\ref{ICineq5})
follows from the fact that $T^{'}$ is a deterministic function of $T$ and invoking the Markov chain in (\ref{ICTP2}). Similarly, we also have
\begin{align}
I(X_{2G};Y_{1},Y_{2},Y_{F_{1}}|X_{1G},T^{'}_{G})&\geq I(X_{2};Y_{1},Y_{2},Y_{F_{1}}|X_{1},T)\label{ICineq7}\\
I(X_{1G},X_{2G};Y_{1},Y_{2},Y_{F_{1}},Y_{F_{2}}|T^{'}_{G})&\geq I(X_{1},X_{2};Y_{1},Y_{2},Y_{F_{1}},Y_{F_{2}}|T)\label{ICineq8}
\end{align}
Finally, we have
\begin{align}
I(X_{1G},X_{2G};Y_{1})&=h(Y_{1})-h(Y_{1}|X_{1G},X_{2G})\label{ICineq9}\\
&=h(X_{1G}+\sqrt{b}X_{2G}+N_{1})-h(N_{1})\label{ICineq10}\\
&\geq h(X_{1}+\sqrt{b}X_{2}+N_{1})-h(N_{1})\label{ICineq11}\\
&=I(X_{1},X_{2};Y_{1})\label{ICineq13}
\end{align}
and similarly, we also have,
\begin{align}
I(X_{1G},X_{2G};Y_{2})&\geq I(X_{1},X_{2};Y_{2})\label{ICineq14}\\
I(X_{1G},X_{2G};Y_{1},Y_{2})&\geq I(X_{1},X_{2};Y_{1},Y_{2})\label{ICineq15}
\end{align}

Therefore, we conclude that for any non-Gaussian distribution
$p(t,x_{1},x_{2})\in \mathcal{P}_{NG}^{DB(b)}$, there exists a
jointly Gaussian distribution $p(t,x_{1},x_{2})\in
\mathcal{P}_{G}^{DB}$ which satisfies the dependence balance bound
(\ref{DBICT8}) and yields a set of rates which includes the set of
rates given by the fixed non-Gaussian distribution. Hence, it
suffices to consider jointly Gaussian distributions in
$\mathcal{P}_{G}^{DB}$ to evaluate our outer bound.

The dependence balance based outer bound can now be written in an explicit form as,
\begin{align}
\mathcal{DB}_{UC}^{IC}=\bigcup_{(\rho_{1T},\rho_{2T}) \in [0,1]\times
[0,1]}\Bigg\{(R_{1},R_{2}): \hspace{0.05in}&R_{1}\leq
\frac{1}{2}\mbox{log}\left(1+f_{1}(\rho_{1T},\rho_{2T})\right)\nonumber\\
&R_{2}\leq
\frac{1}{2}\mbox{log}\left(1+f_{2}(\rho_{1T},\rho_{2T})\right)\nonumber\\
&R_{1}\leq
\frac{1}{2}\mbox{log}\left(1+f_{3}(\rho_{1T})\right)\nonumber\\
&R_{2}\leq \frac{1}{2}\mbox{log}\left(1+f_{4}(\rho_{2T})\right)\nonumber\\
&R_{1}+R_{2}\leq \frac{1}{2}\mbox{log}\left(1+f_{5}(\rho_{1T},\rho_{2T})\right)\nonumber\\
&R_{1}+R_{2}\leq \frac{1}{2}\mbox{log}\left(1+f_{6}(\rho_{1T},\rho_{2T})\right)
\Bigg\}
\end{align}
where
\begin{align}
f_{1}(\rho_{1T},\rho_{2T})&=\frac{(P_{1}+bP_{2}+2\rho_{1T}\rho_{2T}\sqrt{bP_{1}P_{2}})}{\sigma_{N_{1}}^{2}}\\
f_{2}(\rho_{1T},\rho_{2T})&=\frac{(aP_{1}+P_{2}+2\rho_{1T}\rho_{2T}\sqrt{aP_{1}P_{2}})}{\sigma_{N_{2}}^{2}}\\
f_{3}(\rho_{1T})&=(1-\rho_{1T}^{2})P_{1}\left(\frac{1}{\sigma_{N_{1}}^{2}}+\frac{a}{\sigma_{N_{2}}^{2}}+\frac{h_{12}}{\sigma_{Z_{2}}^{2}}\right)
\end{align}
\begin{align}
f_{4}(\rho_{2T})&=(1-\rho_{2T}^{2})P_{2}\left(\frac{b}{\sigma_{N_{1}}^{2}}+\frac{1}{\sigma_{N_{2}}^{2}}+\frac{h_{21}}{\sigma_{Z_{1}}^{2}}\right)\\
f_{5}(\rho_{1T},\rho_{2T})&=f_{1}(\rho_{1T},\rho_{2T})+f_{2}(\rho_{1T},\rho_{2T})+\frac{(1-\rho_{1T}^{2}\rho_{2T}^{2})P_{1}P_{2}(1-\sqrt{ab})^{2}}{\sigma_{N_{1}}^{2}\sigma_{N_{2}}^{2}}\\
f_{6}(\rho_{1T},\rho_{2T})&=f_{3}(\rho_{1T})+f_{4}(\rho_{2T})+(1-\rho_{1T}^{2})(1-\rho_{2T}^{2})P_{1}P_{2}\beta
\end{align}
where
\begin{align}
\beta&=\frac{h_{12}h_{21}}{\sigma_{Z_{1}}^{2}\sigma_{Z_{2}}^{2}} +
\frac{(1-\sqrt{ab})^{2}}{\sigma_{N_{1}}^{2}\sigma_{N_{2}}^{2}} +
\frac{h_{12}}{\sigma_{Z_{2}}^{2}}\left(\frac{1}{\sigma_{N_{2}}^{2}}+\frac{b}{\sigma_{N_{1}}^{2}}\right)
+  \frac{h_{21}}{\sigma_{Z_{1}}^{2}}\left(\frac{1}{\sigma_{N_{1}}^{2}}+\frac{a}{\sigma_{N_{2}}^{2}}\right)
\end{align}

The cut-set outer bound given in (\ref{CSIC1})-(\ref{CSIC6}) is evaluated for the Gaussian IC with user
cooperation described in (\ref{ICUCmodel1})-(\ref{ICUCmodel4}) as
\begin{align}
\mathcal{CS}_{UC}^{IC}=\bigcup_{\rho \in [0,1]}\Bigg\{(R_{1},R_{2}):\hspace{0.05in}
&R_{1}\leq \frac{1}{2}\mbox{log}\left(1+\frac{P_{1}+bP_{2}+2\rho\sqrt{bP_{1}P_{2}}}{\sigma_{N_{1}}^{2}}\right)\nonumber\\
&R_{2}\leq \frac{1}{2}\mbox{log}\left(1+\frac{aP_{1}+P_{2}+2\rho\sqrt{aP_{1}P_{2}}}{\sigma_{N_{2}}^{2}}\right)\nonumber\\
&R_{1}\leq \frac{1}{2}\mbox{log}\left(1+(1-\rho^{2})P_{1}\left(\frac{1}{\sigma_{N_{1}}^{2}}+\frac{a}{\sigma_{N_{2}}^{2}}+\frac{h_{12}}{\sigma_{Z_{2}}^{2}}\right)\right)\nonumber\\
&R_{2}\leq \frac{1}{2}\mbox{log}\left(1+(1-\rho^{2})P_{2}\left(\frac{b}{\sigma_{N_{1}}^{2}}+\frac{1}{\sigma_{N_{2}}^{2}}+\frac{h_{21}}{\sigma_{Z_{1}}^{2}}\right)\right)\nonumber\\
&R_{1}+R_{2}\leq \frac{1}{2}\mbox{log}\left(1+k_{1}(\rho)+k_{2}(\rho)+\frac{(1-\rho^{2})P_{1}P_{2}(1-\sqrt{ab})^{2}}{\sigma_{N_{1}}^{2}\sigma_{N_{2}}^{2}}\right)
\Bigg\}
\end{align}
where
\begin{align}
k_{1}(\rho)&=\frac{P_{1}+bP_{2}+2\rho\sqrt{bP_{1}P_{2}}}{\sigma_{N_{1}}^{2}}\\
k_{1}(\rho)&=\frac{aP_{1}+P_{2}+2\rho\sqrt{aP_{1}P_{2}}}{\sigma_{N_{2}}^{2}}
\end{align}

Figure $13$ illustrates our outer bound, cut-set bound, an achievable rate region with cooperation \cite{Host-Madsen:2006}, capacity region
without cooperation \cite{Sato:1981} for the case when $P_{1}=P_{2}=
\sigma_{N_{1}}^{2}=\sigma_{N_{1}}^{2}=1$ and $\sigma_{Z_{1}}^{2}=\sigma_{Z_{2}}^{2}=1$
and $a=b=1$ and $h_{12}=h_{21}=2$. Figure $14$ illustrates the
outer bound, cut-set bound and achievable region without cooperation \cite{Sason:2004} when
$P_{1}=P_{2}=\sigma_{N_{1}}^{2}=\sigma_{N_{1}}^{2}=1$ and $\sigma_{Z_{1}}^{2}=\sigma_{Z_{2}}^{2}=1$
and $a=b=0.5$ and $h_{12}=h_{21}=0.1$. Figure $15$ illustrates our sum rate upper bound and the cut-set bound as function of $h$, where
 $h=h_{12}=h_{21}$ and $P_{1}=P_{2}=\sigma_{N_{1}}^{2}=\sigma_{N_{1}}^{2}=1$ and $\sigma_{Z_{1}}^{2}=\sigma_{Z_{2}}^{2}=1$, $a=b=0.5$.

 \begin{figure}[p]
   \centerline{\epsfig{figure=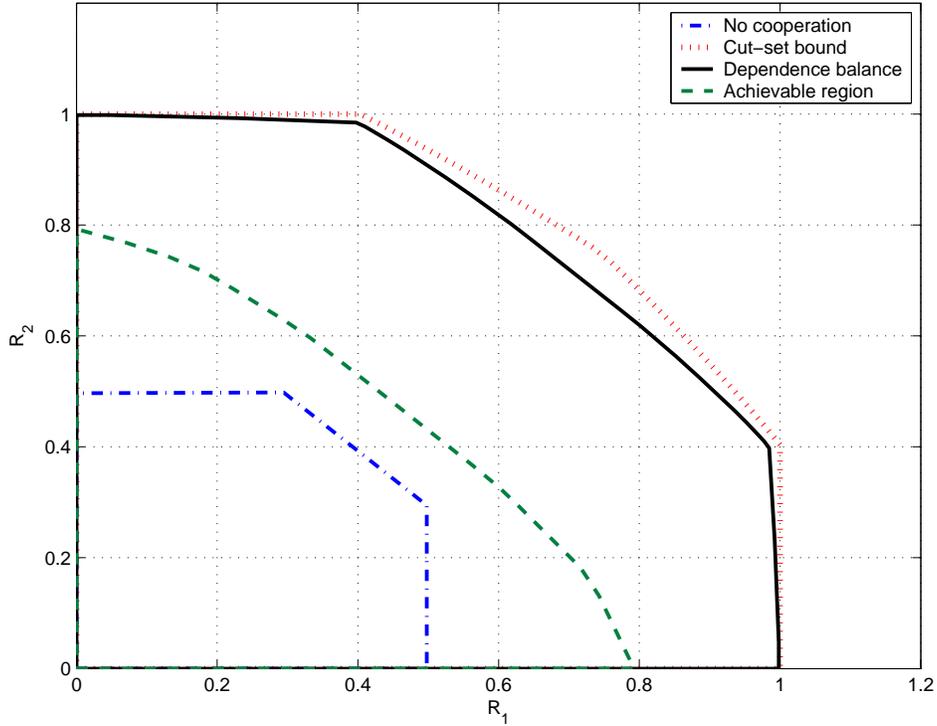,width=12.5cm}}
   \caption{Illustration of bounds for
   $P_{1}=P_{2}=\sigma_{N_{1}}^{2}=\sigma_{N_{2}}^{2}=1$, $\sigma_{Z_{1}}^{2}=\sigma_{Z_{2}}^{2}=1$ and $a=b=1$
    and $h_{12}=h_{21}=2$.}\label{fig13}
   \end{figure}
 \begin{figure}
   \centerline{\epsfig{figure=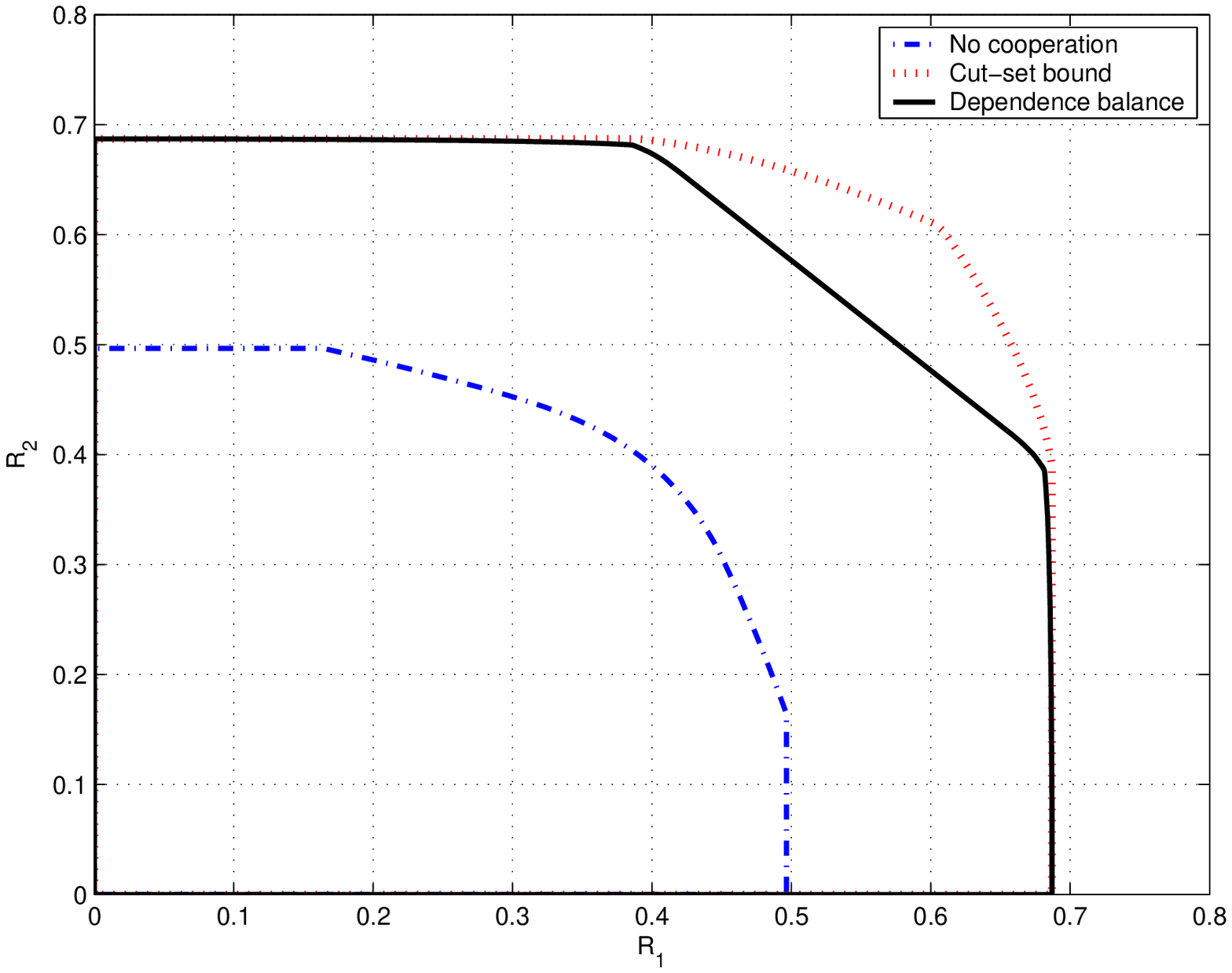,width=12.5cm}}
   \caption{Illustration of bounds for
   for $P_{1}=P_{2}=\sigma_{N_{1}}^{2}=\sigma_{N_{2}}^{2}=1$, $\sigma_{Z_{1}}^{2}=\sigma_{Z_{2}}^{2}=1$ and $a=b=0.5$
    and $h_{12}=h_{21}=0.1$.}\label{fig14}
 \end{figure}
\begin{figure}
  \centerline{\epsfig{figure=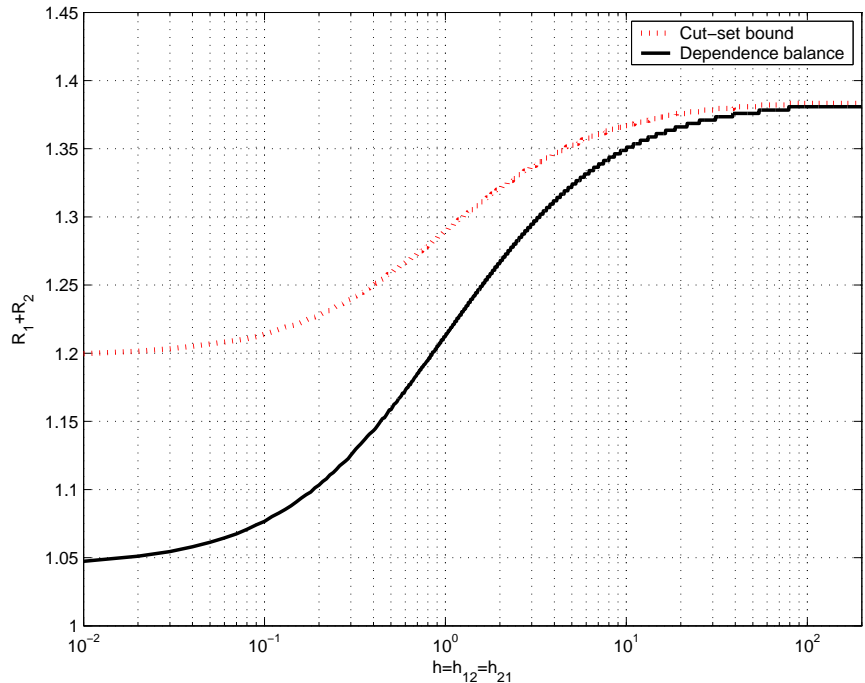,width=11.5cm}}
  \caption{Illustration of sum-rate upper bound and the cut-set bound as a function of $h$, where
    $h=h_{12}=h_{21}$.}\label{fig15}
\end{figure}
\section{Conclusions}
We obtained new outer bounds for the capacity regions of the
two-user MAC with generalized feedback and the two-user IC with
generalized feedback. We explicitly evaluated these outer bounds
for three channel models. In particular, we evaluated our outer
bounds for the Gaussian MAC with different noisy feedback signals
at the transmitters, the Gaussian MAC with user cooperation and
the Gaussian IC with user cooperation. Our outer bounds strictly
improve upon the cut-set bound for all three channel models.

For the evaluation of our outer bounds for the Gaussian scenarios
of interest, we proposed a systematic approach to deal with
capacity bounds involving auxiliary random variables. This
approach was appropriately tailored according to the channel model
in consideration which permitted us to obtain explicit expressions
for our outer bounds. To evaluate our outer bounds, we have to
consider all input distributions satisfying the dependence balance
constraint. The main difficulty in evaluating our outer bounds
arises from the fact that there might exist some non-Gaussian
input distribution $p(t,x_{1},x_{2})$ with a covariance matrix
$Q$, such that $p(t,x_{1},x_{2})$ satisfies the dependence balance
constraint but there does not exist a jointly Gaussian triple with
the covariance matrix $Q$ satisfying the dependence balance
constraint. Therefore, the regular methodology of evaluating outer
bounds, i.e., the approach of applying maximum entropy theorem
\cite{Cover:book} fails beyond this particular point. Through our
explicit evaluation for all three channel models, we were able to
show the existence of a jointly Gaussian triple with a covariance
matrix $S$ which satisfies the dependence balance constraint and
yields larger rates than the fixed non-Gaussian distribution.

In particular, for the case of Gaussian MAC with noisy feedback,
we made use of a recently discovered multivariate EPI
\cite{Palomar:2008}, which is a generalization of Costa's EPI
\cite{Costa:EPI1985}. It is worth nothing that this result could
not be obtained from the classical vector EPI. For the case of
Gaussian MAC with user cooperation and the Gaussian IC with user
cooperation, our proof closely follows a recent result of Bross,
Wigger and Lapidoth \cite{LapidothISIT:2008} and
\cite{Venkat:thesis} for the Gaussian MAC with conferencing
encoders.

\section{Appendix}
\subsection{Proof of Theorem 1}
We will prove Theorem $1$ by first deriving an upper bound for $R_{1}$ as
\begin{align}
nR_{1}&=H(W_{1})=H(W_{1}|W_{2})\\
&=I(W_{1};Y^{n},Y_{F_{2}}^{n}|W_{2})+H(W_{1}|W_{2},Y^{n},Y_{F_{2}}^{n})\\
&\leq I(W_{1};Y^{n},Y_{F_{2}}^{n}|W_{2})+n\epsilon_{1}^{(n)}\label{fano}\\
&=\sum_{i=1}^{n}I(W_{1};Y_{i},Y_{F_{2}i}|W_{2},Y^{i-1},Y_{F_{2}}^{i-1})+n\epsilon_{1}^{(n)}\\
&=\sum_{i=1}^{n}(H(Y_{i},Y_{F_{2}i}|W_{2},Y^{i-1},Y_{F_{2}}^{i-1})-H(Y_{i},Y_{F_{2}i}|W_{1},W_{2},Y^{i-1},Y_{F_{2}}^{i-1}))+n\epsilon_{1}^{(n)}\\
&=\sum_{i=1}^{n}(H(Y_{i},Y_{F_{2}i}|W_{2},X_{2i},Y^{i-1},Y_{F_{2}}^{i-1})-H(Y_{i},Y_{F_{2}i}|X_{2i},W_{1},W_{2},Y^{i-1},Y_{F_{2}}^{i-1}))+n\epsilon_{1}^{(n)}\label{thm0}\\
&\leq
\sum_{i=1}^{n}(H(Y_{i},Y_{F_{2}i}|X_{2i},Y_{F_{2}}^{i-1})-H(Y_{i},Y_{F_{2}i}|X_{2i},W_{1},W_{2},Y^{i-1},Y_{F_{2}}^{i-1}))+n\epsilon_{1}^{(n)}\label{thm1}\\
&\leq
\sum_{i=1}^{n}(H(Y_{i},Y_{F_{2}i}|X_{2i},Y_{F_{2}}^{i-1})-H(Y_{i},Y_{F_{2}i}|X_{1i},X_{2i},W_{1},W_{2},Y^{i-1},Y_{F_{2}}^{i-1}))+n\epsilon_{1}^{(n)}\label{thm2}\\
&= \sum_{i=1}^{n}(H(Y_{i},Y_{F_{2}i}|X_{2i},Y_{F_{2}}^{i-1})-H(Y_{i},Y_{F_{2}i}|X_{1i},X_{2i},Y_{F_{2}}^{i-1}))+n\epsilon_{1}^{(n)}\label{thm3}\\
&= \sum_{i=1}^{n}I(X_{1i};Y_{i},Y_{F_{2}i}|X_{2i},Y_{F_{2}}^{i-1})+n\epsilon_{1}^{(n)}\\
&= nI(X_{1Q};Y_{Q},Y_{F_{2}Q}|X_{2Q},Q,Y_{F_{2}}^{Q-1})+n\epsilon_{1}^{(n)}\\
&= nI(X_{1};Y,Y_{F_{2}}|X_{2},T_{2})+n\epsilon_{1}^{(n)}
\end{align}
where (\ref{fano}) follows from Fano's inequality
\cite{Cover:book}, (\ref{thm0}) follows from the fact that
$X_{2i}$ is a function of $(W_{2},Y_{F_{2}}^{i-1})$ and by
introducing $X_{2i}$ in both terms, (\ref{thm1}) follows from the
fact that conditioning reduces entropy and we drop
$(W_{2},Y^{i-1})$ from the conditioning in the first term,
(\ref{thm2}) follows from the fact that conditioning reduces
entropy and by introducing $X_{1i}$ in the second term and
(\ref{thm3}) follows from the memoryless property of the channel.
Finally, we define $X_{1}=X_{1Q}$, $X_{2}=X_{2Q}$,
$T_{1}=(Q,Y_{F_{1}}^{Q-1})$, $T_{2}=(Q,Y_{F_{2}}^{Q-1})$,
$Y=Y_{Q}$, $Y_{F_{1}}=Y_{F_{1}Q}$ and $Y_{F_{2}}=Y_{F_{2}Q}$,
where $Q$ is a random variable which is uniformly distributed over
$\{1,\ldots , n\}$ and is independent of all other random
variables.
 Similarly, we have
\begin{align}
R_{2}&\leq I(X_{2};Y,Y_{F_{1}}|X_{1},T_{1})\label{thm11}\\
R_{1}+R_{2}&\leq I(X_{1},X_{2};Y,Y_{F_{1}},Y_{F_{2}}|T_{1},T_{2})\label{thm12}
\end{align}
In addition to (\ref{thm12}), we also have the following sum-rate constraint which also appears in the cut-set outer bound,
\begin{align}
n(R_{1}+R_{2})&=H(W_{1},W_{2})\\
&=I(W_{1},W_{2};Y^{n})+H(W_{1},W_{2}|Y^{n})\\
&\leq I(W_{1},W_{2};Y^{n}) + n\epsilon^{(n)}\\
&=\sum_{i=1}^{n}(H(Y_{i}|Y^{i-1})-H(Y_{i}|W_{1},W_{2},Y^{i-1}))+ n\epsilon^{(n)}\\
&\leq \sum_{i=1}^{n}(H(Y_{i}|Y^{i-1})-H(Y_{i}|X_{1i},X_{2i},W_{1},W_{2},Y^{i-1}))+ n\epsilon^{(n)}\\
&= \sum_{i=1}^{n}(H(Y_{i}|Y^{i-1})-H(Y_{i}|X_{1i},X_{2i}))+ n\epsilon^{(n)}\\
&\leq \sum_{i=1}^{n}(H(Y_{i})-H(Y_{i}|X_{1i},X_{2i}))+ n\epsilon^{(n)}\\
&= \sum_{i=1}^{n}I(X_{1i},X_{2i};Y_{i})+ n\epsilon^{(n)}\\
&= nI(X_{1Q},X_{2Q};Y_{Q}|Q)+ n\epsilon^{(n)}\\
&\leq nI(X_{1Q},X_{2Q};Y_{Q})+ n\epsilon^{(n)}\\
&= nI(X_{1},X_{2};Y)+n\epsilon^{(n)}\label{thm13}
\end{align}
It is necessary to include this seemingly trivial upper bound on the sum-rate. The reason for including this sum-rate
upper bound is that one cannot claim that for any input distribution $p(t_{1},t_{2},x_{1},x_{2})$, we
have $I(X_{1},X_{2};Y,Y_{F_{1}},Y_{F_{2}}|T_{1},T_{2})\leq I(X_{1},X_{2};Y)$. In other words, we cannot claim that the sum-rate bound in (\ref{thm13}) will
always be redundant. Therefore, by including it,  we can make sure that our outer bound is at most equal to the cut-set outer bound but never larger than it.
Although, as we will see in the proof of Theorem $3$ for the case of noisy feedback, the sum-rate upper bound in (\ref{thm13}) will turn out to be redundant.

The proof of the dependence balance constraint in (\ref{DBMAC6}) is along the same
lines as in \cite{Hekstra_Willems:1989} by starting from the
inequality
\begin{align}
0\leq I(W_{1};W_{2}|Y_{F_{1}}^{n},Y_{F_{2}}^{n})-I(W_{1};W_{2})
\end{align}
to arrive at
\begin{align}
I(X_{1};X_{2}|T_{1},T_{2})&\leq I(X_{1};X_{2}|Y_{F_{1}},Y_{F_{2}},T_{1},T_{2})\label{proofdbc}
\end{align}
This completes the proof of Theorem $1$.

\subsection{Proof of Theorem 2}
We will prove Theorem $2$ by first deriving an upper bound for $R_{1}$ as
 \begin{align}
 nR_{1}&=H(W_{1})=H(W_{1}|W_{2})\\
 &=I(W_{1};Y_{1}^{n},Y_{2}^{n},Y_{F_{2}}^{n}|W_{2})+H(W_{1}|W_{2},Y_{1}^{n},Y_{2}^{n},Y_{F_{2}}^{n})\\
 &\leq I(W_{1};Y_{1}^{n},Y_{2}^{n},Y_{F_{2}}^{n}|W_{2})+n\epsilon_{1}^{(n)}\label{ICfano}\\
 &=\sum_{i=1}^{n}I(W_{1};Y_{1i},Y_{2i},Y_{F_{2}i}|W_{2},Y_{1}^{i-1},Y_{2}^{i-1},Y_{F_{2}}^{i-1})+n\epsilon_{1}^{(n)}\\
 &=\sum_{i=1}^{n}(H(Y_{1i},Y_{2i},Y_{F_{2}i}|W_{2},Y_{1}^{i-1},Y_{2}^{i-1},Y_{F_{2}}^{i-1})\nonumber\\&\hspace{0.5in}-H(Y_{1i},Y_{2i},Y_{F_{2}i}|W_{1},W_{2},Y_{1}^{i-1},Y_{2}^{i-1},Y_{F_{2}}^{i-1}))+n\epsilon_{1}^{(n)}\\
 &=\sum_{i=1}^{n}(H(Y_{1i},Y_{2i},Y_{F_{2}i}|W_{2},X_{2i},Y_{1}^{i-1},Y_{2}^{i-1},Y_{F_{2}}^{i-1})\nonumber\\&\hspace{0.5in}-H(Y_{1i},Y_{2i},Y_{F_{2}i}|X_{2i},W_{1},W_{2},Y_{1}^{i-1},Y_{2}^{i-1},Y_{F_{2}}^{i-1}))+n\epsilon_{1}^{(n)}\label{ICthm0}\\
 &\leq
 \sum_{i=1}^{n}(H(Y_{1i},Y_{2i},Y_{F_{2}i}|X_{2i},Y_{F_{2}}^{i-1})\nonumber\\&\hspace{0.5in}-H(Y_{1i},Y_{2i},Y_{F_{2}i}|X_{2i},W_{1},W_{2},Y_{1}^{i-1},Y_{2}^{i-1},Y_{F_{2}}^{i-1}))+n\epsilon_{1}^{(n)}\label{ICthm1}\\
 &\leq
 \sum_{i=1}^{n}(H(Y_{1i},Y_{2i},Y_{F_{2}i}|X_{2i},Y_{F_{2}}^{i-1})\nonumber\\&\hspace{0.5in}-H(Y_{1i},Y_{2i},Y_{F_{2}i}|X_{1i},X_{2i},W_{1},W_{2},Y_{1}^{i-1},Y_{2}^{i-1},Y_{F_{2}}^{i-1}))+n\epsilon_{1}^{(n)}\label{ICthm2}\\
 &= \sum_{i=1}^{n}(H(Y_{1i},Y_{2i},Y_{F_{2}i}|X_{2i},Y_{F_{2}}^{i-1})-H(Y_{1i},Y_{2i},Y_{F_{2}i}|X_{1i},X_{2i},Y_{F_{2}}^{i-1}))+n\epsilon_{1}^{(n)}\label{ICthm3}\\
 &= \sum_{i=1}^{n}I(X_{1i};Y_{1i},Y_{2i},Y_{F_{2}i}|X_{2i},Y_{F_{2}}^{i-1})+n\epsilon_{1}^{(n)}\\
 &= nI(X_{1};Y_{1},Y_{2},Y_{F_{2}}|X_{2},T_{2})+n\epsilon_{1}^{(n)}
 \end{align}
 where (\ref{ICfano}) follows from Fano's inequality \cite{Cover:book}, (\ref{ICthm0})
 follows from the fact that $X_{2i}$ is a function of
 $(W_{2},Y_{F_{2}}^{i-1})$ and by introducing $X_{2i}$ in both
 terms, (\ref{ICthm1}) follows from the fact that conditioning reduces
 entropy and we drop $(W_{2},Y_{1}^{i-1},Y_{2}^{i-1})$ from the conditioning in the
 first term, (\ref{ICthm2}) follows from the fact that conditioning reduces entropy
 and by introducing $X_{1i}$ in the conditioning in the second term and (\ref{ICthm3}) follows from the memoryless property
 of the channel. Finally, we define $X_{1}=X_{1Q}$,
 $X_{2}=X_{2Q}$, $T_{1}=(Q,Y_{F_{1}}^{Q-1})$, $T_{2}=(Q,Y_{F_{2}}^{Q-1})$, $Y_{1}=Y_{1Q}$,
 $Y_{2}=Y_{2Q}$, $Y_{F_{1}}=Y_{F_{1}Q}$ and $Y_{F_{2}}=Y_{F_{2}Q}$, where $Q$ is a random variable
 which is uniformly distributed over $\{1,\ldots , n\}$ and is
 independent of all other random variables.

Similarly, we have
 \begin{align}
 R_{2}&\leq I(X_{2};Y_{1},Y_{2},Y_{F_{1}}|X_{1},T_{1})\\
 R_{1}+R_{2}&\leq I(X_{1},X_{2};Y_{1},Y_{2},Y_{F_{1}},Y_{F_{2}}|T_{1},T_{2})
 \end{align}
 and we also have from the cut-set bound
 \begin{align}
 R_{1}&\leq I(X_{1},X_{2};Y_{1})\\
 R_{2}&\leq I(X_{2},X_{2};Y_{2})\\
 R_{1}+R_{2}&\leq I(X_{1},X_{2};Y_{1},Y_{2})
 \end{align}
The proof of the dependence balance constraint is along the same
lines as in \cite{Hekstra_Willems:1989} by starting from the inequality
\begin{align}
0\leq I(W_{1};W_{2}|Y_{F_{1}}^{n},Y_{F_{2}}^{n})-I(W_{1};W_{2})
\end{align}
to arrive at
\begin{align}
I(X_{1};X_{2}|T_{1},T_{2})&\leq I(X_{1};X_{2}|Y_{F_{1}},Y_{F_{2}},T_{1},T_{2})
\end{align}
This completes the proof of Theorem $2$.

\subsection{Proof of Theorem 3}
For any MAC-GF, with transition probabilities in
the form of (\ref{channelstructureNF}), we will obtain a strengthened
version of Theorem $1$. We start by obtaining an upper bound on
$R_{1}$ as
\begin{align}
nR_{1}&=H(W_{1})=H(W_{1}|W_{2})\\
&=I(W_{1};Y^{n},Y_{F_{1}}^{n},Y_{F_{2}}^{n}|W_{2})+H(W_{1}|W_{2},Y^{n},Y_{F_{1}}^{n},Y_{F_{2}}^{n})\\
&\leq I(W_{1};Y^{n},Y_{F_{1}}^{n},Y_{F_{2}}^{n}|W_{2})+n\epsilon_{1}^{(n)}\label{fano2}\\
&=\sum_{i=1}^{n}I(W_{1};Y_{i},Y_{F_{1}i},Y_{F_{2}i}|W_{2},Y^{i-1},Y_{F_{1}}^{i-1},Y_{F_{2}}^{i-1})+n\epsilon_{1}^{(n)}
\end{align}
\begin{align}
&=\sum_{i=1}^{n}I(W_{1};Y_{i}|W_{2},Y^{i-1},Y_{F_{1}}^{i-1},Y_{F_{2}}^{i-1})+n\epsilon_{1}^{(n)}\label{markovnoisy}\\
&=\sum_{i=1}^{n}(H(Y_{i}|W_{2},Y^{i-1},Y_{F_{1}}^{i-1},Y_{F_{2}}^{i-1})-H(Y_{i}|W_{1},W_{2},Y^{i-1},Y_{F_{1}}^{i-1},Y_{F_{2}}^{i-1}))+n\epsilon_{1}^{(n)}\\
&=\sum_{i=1}^{n}(H(Y_{i}|X_{2i},W_{2},Y^{i-1},Y_{F_{1}}^{i-1},Y_{F_{2}}^{i-1})-H(Y_{i}|X_{2i},W_{1},W_{2},Y^{i-1},Y_{F_{1}}^{i-1},Y_{F_{2}}^{i-1}))+n\epsilon_{1}^{(n)}\label{t20}\\
&\leq\sum_{i=1}^{n}(H(Y_{i}|X_{2i},W_{2},Y^{i-1},Y_{F_{1}}^{i-1},Y_{F_{2}}^{i-1})-H(Y_{i}|X_{1i},X_{2i},W_{1},W_{2},Y^{i-1},Y_{F_{1}}^{i-1},Y_{F_{2}}^{i-1}))+n\epsilon_{1}^{(n)}\label{t21}\\
&=
\sum_{i=1}^{n}(H(Y_{i}|X_{2i},W_{2},Y^{i-1},Y_{F_{1}}^{i-1},Y_{F_{2}}^{i-1})-H(Y_{i}|X_{1i},X_{2i},Y_{F_{1}}^{i-1},Y_{F_{2}}^{i-1}))+n\epsilon_{1}^{(n)}\label{t22}\\
&\leq
\sum_{i=1}^{n}(H(Y_{i}|X_{2i},Y_{F_{1}}^{i-1},Y_{F_{2}}^{i-1})-H(Y_{i}|X_{1i},X_{2i},Y_{F_{1}}^{i-1},Y_{F_{2}}^{i-1}))+n\epsilon_{1}^{(n)}\label{t23}\\
&= \sum_{i=1}^{n}I(X_{1i};Y_{i}|X_{2i},Y_{F_{1}}^{i-1},Y_{F_{2}}^{i-1})+n\epsilon_{1}^{(n)}\\
&= nI(X_{1Q};Y_{Q}|X_{2Q},Q,Y_{F_{1}}^{Q-1},Y_{F_{2}}^{Q-1})+n\epsilon_{1}^{(n)}\\
&= nI(X_{1};Y|X_{2},T)+n\epsilon_{1}^{(n)}\label{t24}
\end{align}
where (\ref{fano2}) follows from Fano's inequality \cite{Cover:book},
and (\ref{markovnoisy}) follows from the following Markov chain,
\begin{align}
(Y_{F_{1}i},Y_{F_{2}i})\rightarrow Y_{i} \rightarrow
(W_{1},W_{2},Y^{i-1},Y_{F_{1}}^{i-1},Y_{F_{2}}^{i-1})
\end{align}
and (\ref{t20}) follows from the fact that $X_{2i}$ is a function of
$(W_{2},Y_{F_{2}}^{i-1})$, (\ref{t21}) follows from the fact that
conditioning reduces entropy, (\ref{t22}) from the memoryless property
of the channel and (\ref{t23}) follows by dropping $(W_{2},Y^{i-1})$ from
the first term and obtaining an upper bound. We finally arrive at
(\ref{t24}) by defining the auxiliary random variable
$T=(Q,Y_{F_{1}}^{Q-1},Y_{F_{2}}^{Q-1})$, where $Q$ is a random variable which
is uniformly distributed over $\{1,\ldots,n\}$ and is independent of
all other random variables. Similarly, we also have
\begin{align}
R_{2}&\leq I(X_{2};Y|X_{1},T)
\end{align}
and
\begin{align}
R_{1}+R_{2}&\leq I(X_{1},X_{2};Y,Y_{F_{1}},Y_{F_{2}}|T)\label{t25}\\
&=I(X_{1},X_{2};Y|T)\label{t26}
\end{align}
where (\ref{t26}) follows from the Markov chain
$(Y_{F_{1}},Y_{F_{2}})\rightarrow Y \rightarrow (X_{1},X_{2},T)$.
Moreover, as a consequence of (\ref{t26}), the sum-rate bound
\begin{align}
R_{1}+R_{2}&\leq I(X_{1},X_{2};Y)
\end{align}
obtained in (\ref{thm13}) is redundant for any MAC-GF with transition probabilities in the form of (\ref{channelstructureNF}).
The proof of the dependence balance constraint is the same as in Theorem $1$.
This completes the proof of Theorem $3$.

\subsection{Proof of Theorem 4}
The main idea behind the strengthening of Theorem $1$ for user
cooperation is to use the special conditional probability
structure of (\ref{channelstructure}). Using this conditional
structure, we will obtain an outer bound involving only one
auxiliary random variable. We first note that without any loss of
generality, the conditional distributions $p(y_{F_{1}i}|x_{2i})$
and $p(y_{F_{2}i}|x_{1i})$ can be alternatively expressed as two
deterministic functions \cite{Kramer:2003}, \cite{Willems:1985},
i.e.,
\begin{align}
Y_{F_{1}i}=g_{1}(X_{2i},Z_{1i})\label{chstruc1}\\
Y_{F_{2}i}=g_{2}(X_{1i},Z_{2i})\label{chstruc2}
\end{align}
where the random variables $Z_{1i}$ and $Z_{2i}$ are independent and
identically distributed for all $i\in \{1,\ldots,n\}$ and are also independent of the messages $(W_{1},W_{2})$.
We now prove Theorem $4$ by first obtaining an upper bound on $R_{1}$ as follows,
\begin{align}
nR_{1}&=H(W_{1})=H(W_{1}|W_{2})\\
&=I(W_{1};Y^{n},Y_{F_{2}}^{n},Z_{1}^{n}|W_{2})+H(W_{1}|W_{2},Y^{n},Y_{F_{2}}^{n},Z_{1}^{n})\\
&\leq I(W_{1};Y^{n},Y_{F_{2}}^{n},Z_{1}^{n}|W_{2})+n\epsilon_{1}^{(n)}\label{fano3}\\
&= I(W_{1};Z_{1}^{n}|W_{2})+I(W_{1};Y^{n},Y_{F_{2}}^{n}|W_{2},Z_{1}^{n})+n\epsilon_{1}^{(n)}\label{thm31}\\
&=
I(W_{1};Y^{n},Y_{F_{2}}^{n}|W_{2},Z_{1}^{n})+n\epsilon_{1}^{(n)}\label{thm32}\\
&=\sum_{i=1}^{n}I(W_{1};Y_{i},Y_{F_{2}i}|W_{2},Y^{i-1},Y_{F_{2}}^{i-1},Z_{1}^{n})+n\epsilon_{1}^{(n)}\\
&=\sum_{i=1}^{n}(H(Y_{i},Y_{F_{2}i}|W_{2},Y^{i-1},Y_{F_{2}}^{i-1},Z_{1}^{n})-H(Y_{i},Y_{F_{2}i}|W_{1},W_{2},Y^{i-1},Y_{F_{2}}^{i-1},Z_{1}^{n}))+n\epsilon_{1}^{(n)}\\
&=\sum_{i=1}^{n}(H(Y_{i},Y_{F_{2}i}|W_{2},X_{2i},X_{2}^{i-1},Z_{1}^{n},Y^{i-1},Y_{F_{2}}^{i-1})\nonumber\\&\hspace{0.5in}-H(Y_{i},Y_{F_{2}i}|W_{1},W_{2},Y^{i-1},Y_{F_{2}}^{i-1},Z_{1}^{n}))+n\epsilon_{1}^{(n)}\label{thm33}
\end{align}
\begin{align}
&\leq \sum_{i=1}^{n}(H(Y_{i},Y_{F_{2}i}|W_{2},X_{2i},X_{2}^{i-1},Z_{1}^{n},Y^{i-1},Y_{F_{2}}^{i-1})\nonumber\\&\hspace{0.5in}-H(Y_{i},Y_{F_{2}i}|X_{1i},X_{2i},Y_{F_{1}}^{i-1},Y_{F_{2}}^{i-1},W_{1},W_{2},Y^{i-1},Z_{1}^{n}))+n\epsilon_{1}^{(n)}\label{thm34}\\
&=\sum_{i=1}^{n}(H(Y_{i},Y_{F_{2}i}|W_{2},X_{2i},X_{2}^{i-1},Z_{1}^{n},Y^{i-1},Y_{F_{2}}^{i-1})\nonumber\\&\hspace{0.5in}-H(Y_{i},Y_{F_{2}i}|X_{1i},X_{2i},Y_{F_{1}}^{i-1},Y_{F_{2}}^{i-1}))+n\epsilon_{1}^{(n)}\label{thm35}\\
&=
\sum_{i=1}^{n}(H(Y_{i},Y_{F_{2}i}|X_{2i},X_{2}^{i-1},Y_{F_{1}}^{i-1},Y_{F_{2}}^{i-1},Y^{i-1},W_{2},Z_{1}^{n})\nonumber\\&\hspace{0.5in}-H(Y_{i},Y_{F_{2}i}|X_{1i},X_{2i},Y_{F_{1}}^{i-1},Y_{F_{2}}^{i-1}))+n\epsilon_{1}^{(n)}\label{thm36}\\
&\leq
\sum_{i=1}^{n}(H(Y_{i},Y_{F_{2}i}|X_{2i},Y_{F_{1}}^{i-1},Y_{F_{2}}^{i-1})-H(Y_{i},Y_{F_{2}i}|X_{1i},X_{2i},Y_{F_{1}}^{i-1},Y_{F_{2}}^{i-1}))+n\epsilon_{1}^{(n)}\label{thm37}\\
&= \sum_{i=1}^{n}I(X_{1i};Y_{i},Y_{F_{2}i}|X_{2i},Y_{F_{1}}^{i-1},Y_{F_{2}}^{i-1})+n\epsilon_{1}^{(n)}\\
&= nI(X_{1Q};Y_{Q},Y_{F_{2}Q}|X_{2Q},Q,Y_{F_{1}}^{Q-1},Y_{F_{2}}^{Q-1})+n\epsilon_{1}^{(n)}\\
&= nI(X_{1};Y,Y_{F_{2}}|X_{2},T)+n\epsilon_{1}^{(n)}
\end{align}
where (\ref{fano3}) follows from Fano's inequality
\cite{Cover:book}, (\ref{thm32}) follows from the independence of
$(W_{1},W_{2})$ and $Z_{1}^{n}$, (\ref{thm33}) follows by adding
$(X_{2i},X_{2}^{i-1})$ in the conditioning of the first term. This
is possible since $(X_{2i},X_{2}^{i-1})$ is a function of
$(W_{2},Y_{F_{2}}^{i-1})$. We further upper bound by introducing
$(X_{1i},X_{2i},Y_{F_{1}}^{i-1})$ in the conditioning in the
second term to arrive at (\ref{thm34}). In (\ref{thm35}), we use
the memoryless property of the channel to drop
$(W_{1},W_{2},Y^{i-1},Z_{1}^{n})$ from the conditioning in the
second term while retaining $(Y_{F_{1}}^{i-1},Y_{F_{2}}^{i-1})$.

Next, we make use of the special channel structure of (\ref{channelstructure}). More specifically,  using (\ref{chstruc1}), we observe that
 $Y_{F_{1}}^{i-1}$ is a deterministic function of $X_{2}^{i-1}$ and $Z_{1}^{i-1}$ and therefore, it is introduced in the conditioning in the first term in (\ref{thm36}).
This is the crucial part of the proof which enables us to obtain an outer bound involving only one auxiliary random variable as opposed to two auxiliary random variables.
Next, we upper bound (\ref{thm36}) by dropping
$(W_{2},Y^{i-1},X_{2}^{i-1},Z_{1}^{n})$ from the first term to arrive at (\ref{thm37}).
Finally, we define $T=(Q,Y_{F_{1}}^{Q-1},Y_{F_{2}}^{Q-1})$, $X_{1}=X_{1Q}$,
$X_{2}=X_{2Q}$, $Y=Y_{Q}$, $Y_{F_{1}}=Y_{F_{1}Q}$ and $Y_{F_{2}}=Y_{F_{2}Q}$, where $Q$ is a
random variable which is uniformly distributed over $\{1,\ldots,n\}$ and
is independent of all other random variables. Similarly, we have
\begin{align}
R_{2}&\leq I(X_{2};Y,Y_{F_{1}}|X_{1},T)\\
R_{1}+R_{2}&\leq I(X_{1},X_{2};Y,Y_{F_{1}},Y_{F_{2}}|T)
\end{align}
The derivation of the constraint (\ref{DBT4}) is the same as in
Theorem $1$ and is omitted. Moreover, from the proof of the dependence balance
constraint in (\ref{DBMAC6}), we observe that $Y_{F_{1}}^{i-1}$ and $Y_{F_{2}}^{i-1}$ appear together in the
conditioning. Therefore, from our earlier definition of $T=(Q,Y_{F_{1}}^{Q-1},Y_{F_{2}}^{Q-1})$, we directly have from the proof of (\ref{DBMAC6})
\begin{align}
I(X_{1};X_{2}|T)&\leq I(X_{1};X_{2}|Y_{F_{1}},Y_{F_{2}},T)
\end{align}
This completes the proof of Theorem $4$.

\subsection{Proof of Theorem 5}
The idea behind obtaining a strengthened version of Theorem $2$ for IC with user cooperation
is to use the special transition probability structure of (\ref{channelstructureIC}). Using the same argument as in
the proof of Theorem $4$, we can express $Y_{F_{1}i}$ and $Y_{F_{2}i}$ as,
\begin{align}
Y_{F_{1}i}=g_{1}(X_{2i},Z_{1i})\label{ICchstruc1}\\
Y_{F_{2}i}=g_{2}(X_{1i},Z_{2i})\label{ICchstruc2}
\end{align}
where the random variables $Z_{1i}$ and $Z_{2i}$ are independent and
identically distributed for all $i\in \{1,\ldots,n\}$ and are also independent of the messages $(W_{1},W_{2})$.
We now prove Theorem $5$ by first obtaining an upper bound on $R_{1}$ as follows,
\begin{align}
nR_{1}&=H(W_{1})=H(W_{1}|W_{2})\\
&=I(W_{1};Y_{1}^{n},Y_{2}^{n},Y_{F_{2}}^{n},Z_{1}^{n}|W_{2})+H(W_{1}|W_{2},Y_{1}^{n},Y_{2}^{n},Y_{F_{2}}^{n},Z_{1}^{n})\\
&\leq I(W_{1};Y_{1}^{n},Y_{2}^{n},Y_{F_{2}}^{n},Z_{1}^{n}|W_{2})+n\epsilon_{1}^{(n)}\label{ICfano3}\\
&= I(W_{1};Z_{1}^{n}|W_{2})+I(W_{1};Y_{1}^{n},Y_{2}^{n},Y_{F_{2}}^{n}|W_{2},Z_{1}^{n})+n\epsilon_{1}^{(n)}\label{ICthm31}\\
&= I(W_{1};Y_{1}^{n},Y_{2}^{n},Y_{F_{2}}^{n}|W_{2},Z_{1}^{n})+n\epsilon_{1}^{(n)}\label{ICthm32}\\
&=\sum_{i=1}^{n}I(W_{1};Y_{1i},Y_{2i},Y_{F_{2}i}|W_{2},Y_{1}^{i-1},Y_{2}^{i-1},Y_{F_{2}}^{i-1},Z_{1}^{n})+n\epsilon_{1}^{(n)}\\
&=\sum_{i=1}^{n}(H(Y_{1i},Y_{2i},Y_{F_{2}i}|W_{2},Y_{1}^{i-1},Y_{2}^{i-1},Y_{F_{2}}^{i-1},Z_{1}^{n})\nonumber\\&\hspace{0.5in}-H(Y_{1i},Y_{2i},Y_{F_{2}i}|W_{1},W_{2},Y_{1}^{i-1},Y_{2}^{i-1},Y_{F_{2}}^{i-1},Z_{1}^{n}))+n\epsilon_{1}^{(n)}\\
&=\sum_{i=1}^{n}(H(Y_{1i},Y_{2i},Y_{F_{2}i}|W_{2},X_{2i},X_{2}^{i-1},Z_{1}^{n},Y_{1}^{i-1},Y_{2}^{i-1},Y_{F_{2}}^{i-1})\nonumber\\&\hspace{0.5in}-H(Y_{1i},Y_{2i},Y_{F_{2}i}|W_{1},W_{2},Y_{1}^{i-1},Y_{2}^{i-1},Y_{F_{2}}^{i-1},Z_{1}^{n}))+n\epsilon_{1}^{(n)}\label{ICthm33}\\
&\leq
\sum_{i=1}^{n}(H(Y_{1i},Y_{2i},Y_{F_{2}i}|W_{2},X_{2i},X_{2}^{i-1},Z_{1}^{n},Y_{1}^{i-1},Y_{2}^{i-1},Y_{F_{2}}^{i-1})\nonumber\\&\hspace{0.5in}-H(Y_{1i},Y_{2i},Y_{F_{2}i}|X_{1i},X_{2i},Y_{F_{1}}^{i-1},Y_{F_{2}}^{i-1},W_{1},W_{2},Y_{1}^{i-1},Y_{2}^{i-1},Z_{1}^{n}))+n\epsilon_{1}^{(n)}\label{ICthm34}
\end{align}
\begin{align}
&=\sum_{i=1}^{n}(H(Y_{1i},Y_{2i},Y_{F_{2}i}|W_{2},X_{2i},X_{2}^{i-1},Z_{1}^{n},Y_{1}^{i-1},Y_{2}^{i-1},Y_{F_{2}}^{i-1})\nonumber\\&\hspace{0.5in}-H(Y_{1i},Y_{2i},Y_{F_{2}i}|X_{1i},X_{2i},Y_{F_{1}}^{i-1},Y_{F_{2}}^{i-1}))+n\epsilon_{1}^{(n)}\label{ICthm35}\\
&=
\sum_{i=1}^{n}(H(Y_{1i},Y_{2i},Y_{F_{2}i}|X_{2i},X_{2}^{i-1},Y_{F_{1}}^{i-1},Y_{F_{2}}^{i-1},Y_{1}^{i-1},Y_{2}^{i-1},W_{2},Z_{1}^{n})\nonumber\\&\hspace{0.5in}-H(Y_{1i},Y_{2i},Y_{F_{2}i}|X_{1i},X_{2i},Y_{F_{1}}^{i-1},Y_{F_{2}}^{i-1}))+n\epsilon_{1}^{(n)}\label{ICthm36}\\
&\leq
\sum_{i=1}^{n}(H(Y_{1i},Y_{2i},Y_{F_{2}i}|X_{2i},Y_{F_{1}}^{i-1},Y_{F_{2}}^{i-1})-H(Y_{1i},Y_{2i},Y_{F_{2}i}|X_{1i},X_{2i},Y_{F_{1}}^{i-1},Y_{F_{2}}^{i-1}))+n\epsilon_{1}^{(n)}\label{ICthm37}\\
&= \sum_{i=1}^{n}I(X_{1i};Y_{1i},Y_{2i},Y_{F_{2}i}|X_{2i},Y_{F_{1}}^{i-1},Y_{F_{2}}^{i-1})+n\epsilon_{1}^{(n)}\\
&= nI(X_{1Q};Y_{1Q},Y_{2Q},Y_{F_{2}Q}|X_{2Q},Q,Y_{F_{1}}^{Q-1},Y_{F_{2}}^{Q-1})+n\epsilon_{1}^{(n)}\\
&= nI(X_{1};Y_{1},Y_{2},Y_{F_{2}}|X_{2},T)+n\epsilon_{1}^{(n)}
\end{align}
where (\ref{ICfano3}) follows from Fano's inequality
\cite{Cover:book}, (\ref{ICthm32}) follows from the independence
of $(W_{1},W_{2})$ and $Z_{1}^{n}$, (\ref{ICthm33}) follows by
adding $(X_{2i},X_{2}^{i-1})$ in the conditioning of the first
term. This is possible since $(X_{2i},X_{2}^{i-1})$ is a function
of $(W_{2},Y_{F_{2}}^{i-1})$. We further upper bound by
introducing $(X_{1i},X_{2i},Y_{F_{1}}^{i-1})$ in the conditioning
in the second term to arrive at (\ref{ICthm34}). In
(\ref{ICthm35}), we use the memoryless property of the channel to
drop $(W_{1},W_{2},Y_{1}^{i-1},Y_{2}^{i-1},Z_{1}^{n})$ from the
conditioning in the second term while retaining
$(Y_{F_{1}}^{i-1},Y_{F_{2}}^{i-1})$.

Next, we make use of the special channel structure of (\ref{channelstructureIC}). More specifically,  using (\ref{ICchstruc1}), we observe that
 $Y_{F_{1}}^{i-1}$ is a deterministic function of $X_{2}^{i-1}$ and $Z_{1}^{i-1}$ and therefore, it is introduced in the conditioning in the first term in (\ref{ICthm36}).
Next, we upper bound (\ref{ICthm36}) by dropping
$(W_{2},X_{2}^{i-1},Y_{1}^{i-1},Y_{2}^{i-1},Z_{1}^{n})$ from the first term to arrive at (\ref{ICthm37}).
Finally, we define $T=(Q,Y_{F_{1}}^{Q-1},Y_{F_{2}}^{Q-1})$, $X_{1}=X_{1Q}$,
$X_{2}=X_{2Q}$, $Y_{1}=Y_{1Q}$, $Y_{2}=Y_{2Q}$, $Y_{F_{1}}=Y_{F_{1}Q}$ and $Y_{F_{2}}=Y_{F_{2}Q}$, where $Q$ is a
random variable which is uniformly distributed over $\{1,\ldots,n\}$ and
is independent of all other random variables. Similarly, we have
 \begin{align}
 R_{2}&\leq I(X_{2};Y_{1},Y_{2},Y_{F_{1}}|X_{1},T)
 \end{align}
 The derivations of the remaining constraints are similar to the proof of Theorem $2$ since
 both $Y_{F_{1}}^{i-1}$ and $Y_{F_{2}}^{i-1}$ appear together in the conditioning and
 $T$ can be defined appropriately without any difficulty. The proof of dependence balance constraint
 in (\ref{DBICT8}) is the same as in Theorem $2$.
This completes the proof of Theorem $5$.

\subsection{Proof of (\ref{E32})}

In the following derivation of (\ref{E32}), we have dropped conditioning on
$T=t$, for the purpose of simplicity. Substituting
(\ref{lambdadef}), (\ref{kappadef}), (\ref{mudef}) and
(\ref{Vdef}) in (\ref{Palomar}), we have
\begin{align}
N(\Lambda_{1}^{1/2}\mathbf{Y}+\mathbf{V})&=N(\sqrt{\kappa}Y+V_{1},V_{2})\\
&=\frac{1}{(2\pi\mbox{e})}\mbox{e}^{h(\sqrt{\kappa}Y+V_{1},V_{2})}\\
&=
\frac{1}{\sqrt{(2\pi\mbox{e})}}\mbox{e}^{h(\sqrt{\kappa}Y+V_{1})}\label{lamb1}
\end{align}
We also note the following inequality,
\begin{align}
h(\sqrt{\kappa}Y+V_{1}) &\geq \frac{1}{2}\mbox{log}(\mbox{e}^{2h(\sqrt{\kappa}Y)}+2\pi\mbox{e})\label{lamb2}\\
&= \frac{1}{2}\mbox{log}(\kappa
\mbox{e}^{2h(Y)}+2\pi\mbox{e})\label{lamb3}
\end{align}
where (\ref{lamb2}) follows from the scalar EPI \cite{Cover:book}
and (\ref{lamb3}) follows from the fact that for any scalar $c$,
$h(cY)=h(Y)+\mbox{log}(|c|)$ \cite{Cover:book}. Substituting
(\ref{lamb3}) in (\ref{lamb1}), we obtain
\begin{align}
N(\Lambda_{1}^{1/2}\mathbf{Y}+\mathbf{V})&\geq \left(\frac{\kappa
\mbox{e}^{2h(Y)}+2\pi\mbox{e}}{(2\pi\mbox{e})}\right)^{1/2}\label{lamb4}
\end{align}
Similarly, we also have
\begin{align}
N(\Lambda_{2}^{1/2}\mathbf{Y}+\mathbf{V})&\geq \left(\frac{\kappa
\mbox{e}^{2h(Y)}+2\pi\mbox{e}}{(2\pi\mbox{e})}\right)^{1/2}\label{lamb5}
\end{align}
Therefore, we have
\begin{align}
\mu N(\Lambda_{1}^{1/2}\mathbf{Y}+\mathbf{V})+(1-\mu)
N(\Lambda_{2}^{1/2}\mathbf{Y}+\mathbf{V})&\geq \left(\frac{\kappa
\mbox{e}^{2h(Y)}+2\pi\mbox{e}}{(2\pi\mbox{e})}\right)^{1/2}\label{lamb6}
\end{align}
Moreover, the right hand side of (\ref{Palomar}) simplifies to,
\begin{align}
N((\mu\Lambda_{1}+(1-\mu)\Lambda_{2})^{1/2}\mathbf{Y}+\mathbf{V})&=\frac{1}{(2\pi\mbox{e})}\mbox{e}^{h(\sqrt{\mu\kappa}Y+V_{1},\sqrt{(1-\mu)\kappa}Y+V_{2})}\\
&=\frac{1}{(2\pi\mbox{e})}\mbox{e}^{h((\mu\Lambda_{1}+(1-\mu)\Lambda_{2})^{1/2}[Y_{F_{1}} \quad Y_{F_{2}}]^{T})}\\
&=\frac{1}{(2\pi\mbox{e})\sqrt{\sigma_{Z_{1}}^{2}\sigma_{Z_{2}}^{2}}}\mbox{e}^{h(Y_{F_{1}},Y_{F_{2}})}\label{lamb7}
\end{align}
Using (\ref{lamb5})-(\ref{lamb7}) and substituting in
(\ref{Palomar}), we obtain
\begin{align}
\left(\frac{\kappa
\mbox{e}^{2h(Y)}+2\pi\mbox{e}}{(2\pi\mbox{e})}\right)^{1/2}&\leq
\frac{1}{(2\pi\mbox{e})\sqrt{\sigma_{Z_{1}}^{2}\sigma_{Z_{2}}^{2}}}\mbox{e}^{h(Y_{F_{1}},Y_{F_{2}})}\label{lamb8}
\end{align}
Simplifying (\ref{lamb8}) by substituting the value of $\kappa$
and reintroducing the conditioning on $T=t$, we have the proof of
(\ref{E32}),
\begin{align}
h(Y_{F_{1}},Y_{F_{2}}|T=t)&\geq \frac{1}{2}\mbox{log}\left((2\pi
\mbox{e})^{2}\sigma_{Z_{1}}^{2}\sigma_{Z_{2}}^{2}+2\pi\mbox{e}(\sigma_{Z_{1}}^{2}+\sigma_{Z_{2}}^{2})\mbox{e}^{2h(Y|T=t)}\right)
\end{align}

\bibliography{bibliographyjournal}
\bibliographystyle{unsrt}
\end{document}